\documentclass[a4paper,11pt]{article}
\pdfoutput=1

\usepackage{jinstpub}
\usepackage[english]{babel}
\usepackage{upgreek}
\usepackage{lineno}
\usepackage{xspace}
\usepackage{graphicx}
\usepackage{booktabs}
\usepackage[capitalize]{cleveref}
\usepackage{enumitem} 
\usepackage[textsize=footnotesize]{todonotes}

\graphicspath{{Figs/}} 

\newcommand{\xmax}[0]{\ensuremath{X_\text{max}}}

\title{\boldmath The Scintillator Surface Detector of the Pierre Auger Observatory}
\date{\today}


\author{\includegraphics[height=30mm]{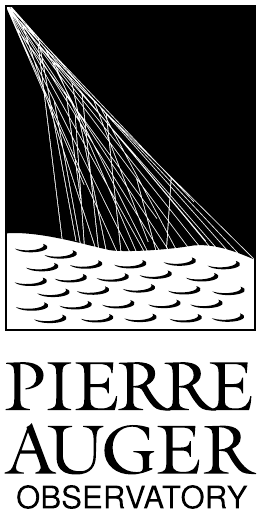}\\[3mm]The Pierre Auger Collaboration}
\affiliation{Av.\ San Mart\'{\i}n Norte 306, 5613 Malarg\"ue, Mendoza, Argentina}

\emailAdd{spokespersons@auger.org}

\abstract{Data collected so far by the Pierre Auger Observatory have enabled major advances in ultra-high energy cosmic ray physics and demonstrated that improved determination of masses of primary cosmic-ray particles, preferably on an event-by-event basis, is necessary for understanding their origin and nature. Improvement in primary mass measurements was the main motivation for the upgrade of the Pierre Auger Observatory, called AugerPrime. As part of this upgrade, scintillator detectors are added to the existing water-Cherenkov surface detector stations. By making use of the differences in detector response to the electromagnetic particles and muons between scintillator and water-Cherenkov detectors, the electromagnetic and muonic components of cosmic-ray air showers can be disentangled. Since the muonic component is sensitive to the primary mass, such combination of detectors provides a powerful way to improve primary mass composition measurements over the original Auger surface detector design. In this paper, the so-called Scintillator Surface Detectors are discussed, including their design characteristics, production process, testing procedure and deployment in the field.}

\begin{document}

\maketitle

\flushbottom

\section{Introduction}
\label{sec:intro}

The Pierre Auger Observatory~\cite{00PAOref} is located on the vast plain known as the Pampa Amarilla near the city of Malarg\"ue, in the province of Mendoza, Argentina.
It is designed to detect ultra-high energy cosmic ray (UHECR) showers using a hybrid combination of two almost independent systems: the surface detector (SD) and the fluorescence detector (FD).
The SD covers an area of about 3000\,km$^2$ and consists of an array of  1660 water-Cherenkov detectors (WCD) arranged on a triangular grid with a separation of 1.5\,km (750 m in a sub-array). It provides lateral sampling of extensive air showers (EAS) at ground level.
The FD is a system of 27 fluorescence telescopes located at four sites on the perimeter of the array observing the atmosphere above the SD. It measures the shower longitudinal profile based on measurement of the UV fluorescence light emitted by the de-excitation of nitrogen molecules excited as the shower passes through the atmosphere.

The large quantity of high-quality data recorded since 2004 with the Pierre Auger Observatory has led to a number of major breakthroughs in the field of UHECRs.
Examples include  precise measurements of the cosmic ray energy spectrum over a wide range of energies, uncovering new features and unambiguously establishing a flux suppression above $4{\times}10^{19}$\,eV~\cite{PierreAuger:2020kuy, PierreAuger:2020qqz}; providing stringent limits on photon \cite{Abreu_2023} and neutrino \cite{Aab_2019} fluxes at ultra-high energy; and identifying directional anisotropies at medium~\cite{PierreAuger:2022axr} and large~\cite{Aab_2018} angular scales.   A surprising finding, at least based on the conventional wisdom at the time of the Observatory's conception, is that the cosmic ray composition at the highest energies is not protonic, but instead becomes heavier with increasing energy above $3{\times}10^{18}$\,eV~\cite{PierreAuger:2014sui}.  This result is mainly based on observations of the depth of shower maximum (\xmax) by the FD with its limited (approximately 13\%) duty cycle constrained by favorable night time atmospheric conditions.

The goal of UHECR physics is the identification of sources (or source classes) responsible for the acceleration of the highest energy cosmic rays. Given the more complex primary composition picture, this goal can only be achieved if the primary mass is estimated on a near event-by-event basis, together with precise measurements of the energy and arrival direction, since the mass and energy determine the bending of the cosmic ray path in galactic and extra-galactic magnetic fields.   

In its original configuration, the Pierre Auger Observatory can provide measurements of the energy and arrival direction of cosmic rays with unprecedented quality, but it has a limited capability for identifying their primary mass on an event-by event basis, particularly at times when the FD is not operating.  However, new capabilities of the surface detector are being pursued that allow for mass estimates for all events.  This is the basis of the AugerPrime upgrade of the Observatory, including the new Scintillator Surface Detectors (SSDs), the subject of this article.
  
\begin{figure}
\centering
\def\h{0.4}
\includegraphics[height=\h\textwidth]{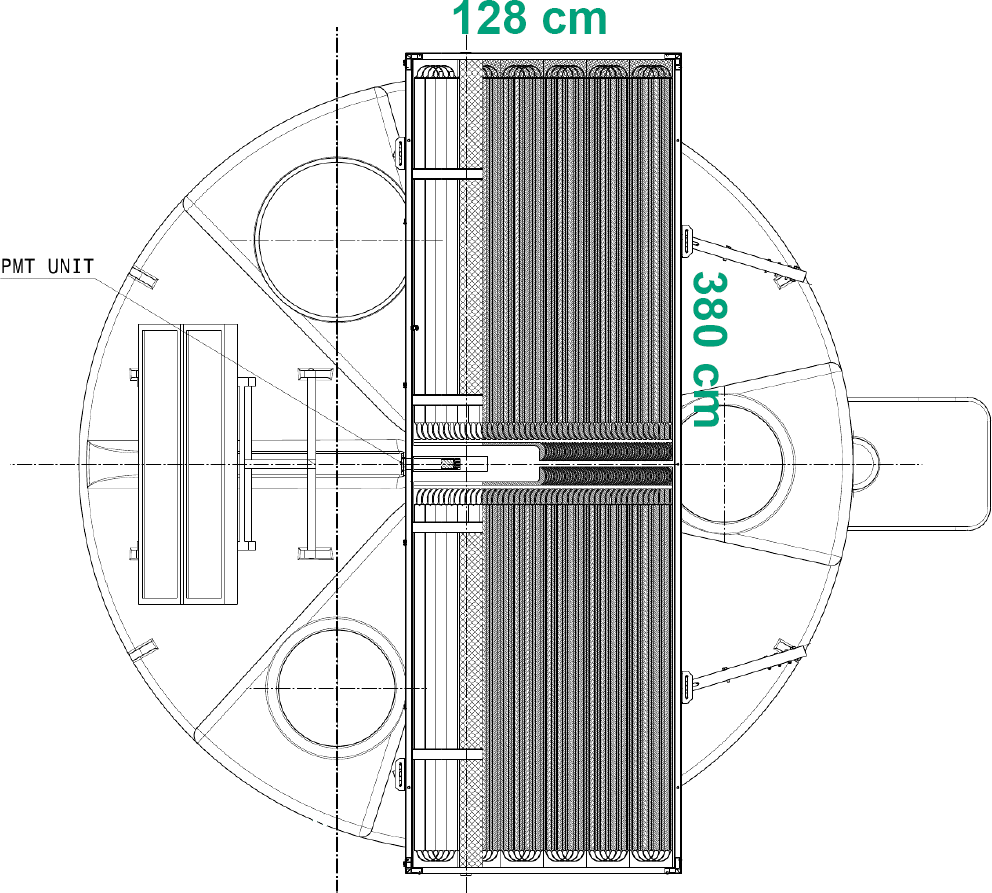}\hfill
\includegraphics[height=\h\textwidth]{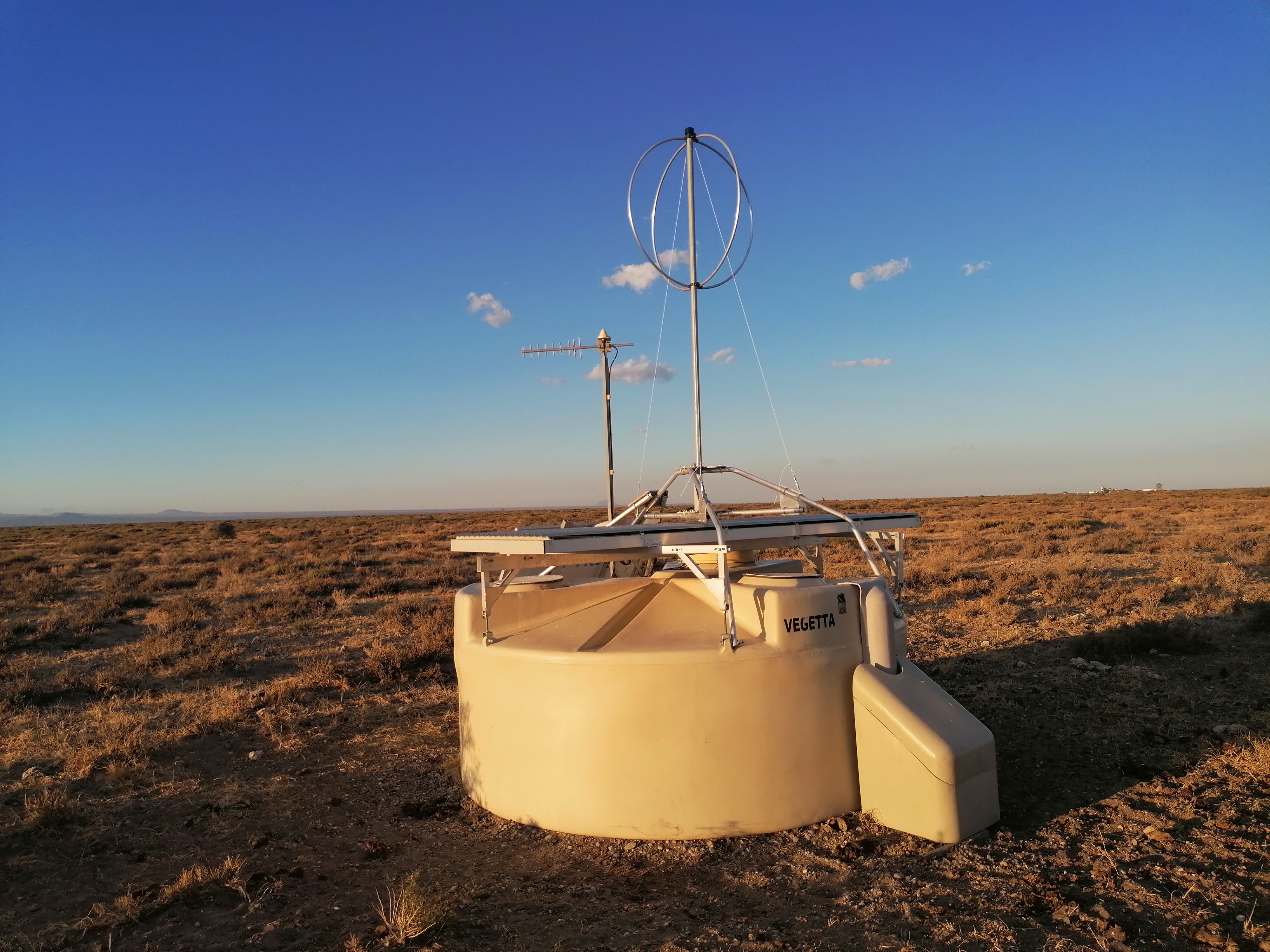}
\caption{\emph{Left:} Drawing of an open scintillator detector.
\emph{Right:} Photograph of one of the stations of the Surface Detector of the Pierre Auger Observatory, after its upgrade.}
\label{fig:fullyUPLS}
\end{figure}

One technique for inferring the mass of the primary cosmic ray is through analysis of the different particle components of the extensive air shower it produces.  While the signals in a WCD are sensitive to muon track length and energy of the electromagnetic particles, a thin plastic scintillator acts as a particle counter, so it is relatively more sensitive to the electromagnetic component (since electrons/gammas are more numerous than muons not too far from the shower core). 
Thus, combined measurements with both water-Cherenkov and scintillator detectors allow disentangling the air-shower muonic and electromagnetic components and greatly assist with primary mass identification. The formalism \cite{ALS2014} developed for layered surface detectors was adapted for this purpose \cite{01intro}. The SSD/WCD combination will be useful for mass studies for showers with zenith angles $<60^\circ$, i.e. showers with a significant electromagnetic component at ground-level and negligible reconstruction bias due to geometry of thin scintillators.
SSD units are now installed on top of each water-Cherenkov detector at the Observatory (\cref{fig:fullyUPLS}), with the exception of the ring of WCDs on the border of the array.  

The  other elements of the AugerPrime upgrade of the Observatory are:
\begin{itemize}
\item a Radio Antenna (RD) located on top of the water-Cherenkov detector~\cite{radio} to provide additional mass composition information, especially for showers detected at larger zenith angles; 
\item an additional Small Photomultiplier (SPMT) installed inside the WCD~\cite{small} to extend the dynamic range of the measurable particle densities (e.g. close to the EAS core);
\item an array of Underground Muon Detectors (UMD) deployed near each WCD within the 750 metre in-fill array ~\cite{amiga} to provide direct measurements of the muon component of air showers. This will enable verification and fine-tuning of methods used to extract muon information from the combined SSD and WCD measurements; 
\item new detector electronics (Upgraded Unified Board, UUB) that processes the measurements collected from all the detectors with improved performance~\cite{UUB}. The use of the UUB enables managing all the new detectors in addition to the existing ones and provides a faster sampling of ADC traces from the detectors and better timing accuracy, thus enhancing the local trigger and processing capabilities.
\end{itemize}

In this paper we will describe the Scintillator Surface Detector design, production, tests and deployment. Section 2 is dedicated to the detailed design of the detector, while in section 3 the adopted production procedures are described. In sections 4 and 5, the quality checks used to verify the detectors and PMTs are outlined. Section 6 is used to describe the deployment procedure, followed by conclusions and a first example of an event collected with the upgraded Observatory.  

\section{Design of the Scintillator Surface Detector}

\begin{figure}
\centering
\def\w{0.49}
\includegraphics[width=\w\textwidth]{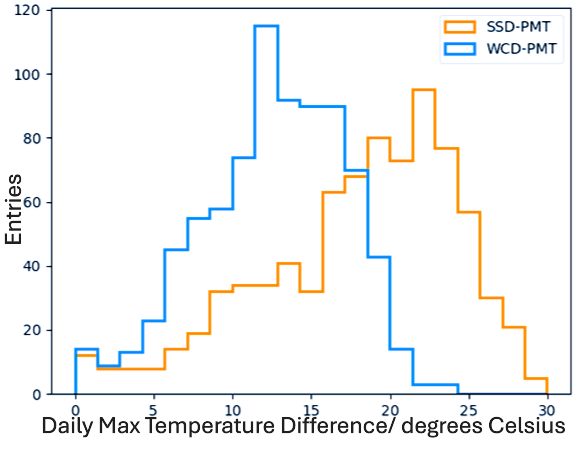}\hfill
\includegraphics[width=\w\textwidth]{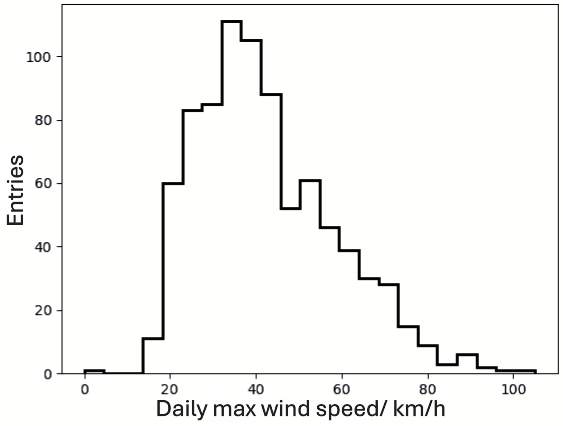}
\caption{\emph{Left:} Diurnal temperature variation measured in the WCD and SSD for one of the stations of the array over a period of two years. The WCD-PMTs suffer smaller temperature differences due to the large mass of water in the tank.
\emph{Right:} Maximum wind speed monitored in the area over the same period of two years.}
\label{fig:tempwind}
\end{figure}

The Pampa Amarilla lies in the rain shadow of the Andes and its climate is classified as arid~\cite{dessur}.
In the experimental area, harsh environmental conditions are experienced.
The ground-level conditions (air pressure, temperature, wind-speed, humidity etc.) are monitored every five minutes by a series of weather stations at each of the four fluorescence detector sites, and near the centre of the array at the Central Laser Facility \cite{00PAOref}.
The recorded temperature range is around $50^\circ$C across the year, and up to $20^\circ-30^\circ$C during the day.   
The graph in \cref{fig:tempwind}-Left shows the diurnal temperature variation over a data-taking period of two years for one of the WCD/SSD stations of the array. The WCD temperature sensor is located on the PMT in the water tank, while the SSD temperature sensor is positioned inside its PMT housing. The difference measured by the two sensors follows expectations since the PMT in the tank is thermally stabilized by the presence of the water.
During the same period, a wind speed of more than 100\,km/h  has been measured (\cref{fig:tempwind}-Right). Such a wind speed imposes the necessity to use firmly anchored structures.

\begin{figure}
\centering
\def\w{0.94}
\includegraphics[width=\textwidth]{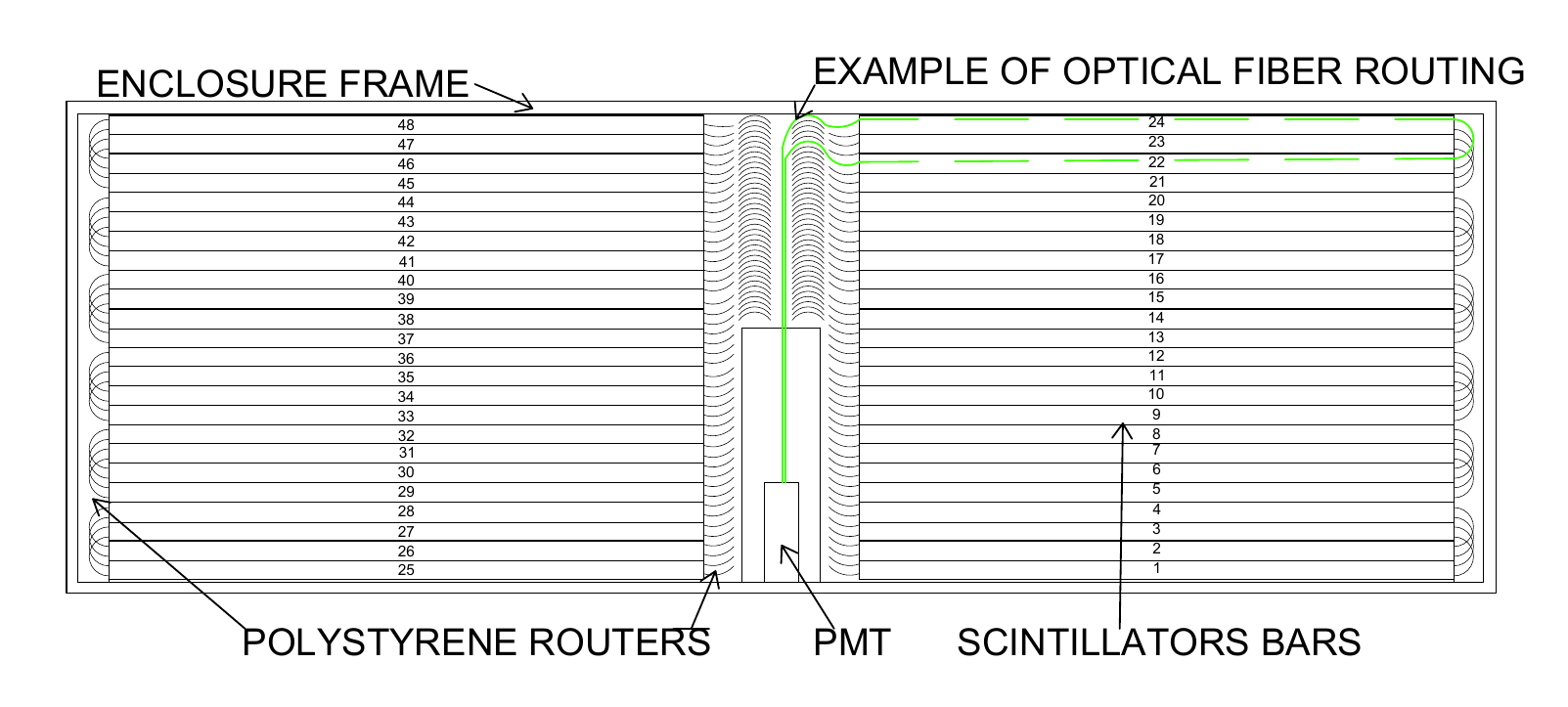}
\caption{Schematic of an open SSD, showing the layout of the components. The green dashed line exemplifies the ``U'' routing of optical fibers, see the text and \cref{fig:ssdscint}.}
\label{fig:schema}
\end{figure}

The SSD module consists of two scintillator panels composed of organic plastic scintillator bars,  encased in an aluminium box, with a photomultiplier (PMT) housed between the panels. The total active area of the scintillators in a module is 3.84\,m$^2$.
A schematic view of an open SSD with its components is shown in \cref{fig:schema}. The module dimensions, 3800\,mm $\times$ 1280\,mm, were chosen to maximize the detector area and overlap with the WCD without impairing the WCD maintenance.

The design of the mechanical structure of the enclosure must guarantee light tightness and robustness for more than ten years of operation, and enough rigidity for transportation, as well as electromagnetic shielding of the PMT.
The remoteness and large extent of the experimental area impose a requirement that only a limited maintenance of the detectors be needed, and for this reason the access to the PMT must be easy and, as much as possible, separated from the area where the scintillator elements are placed.

\begin{figure}
\centering
\def\w{0.95}
\includegraphics[width=\w\textwidth]{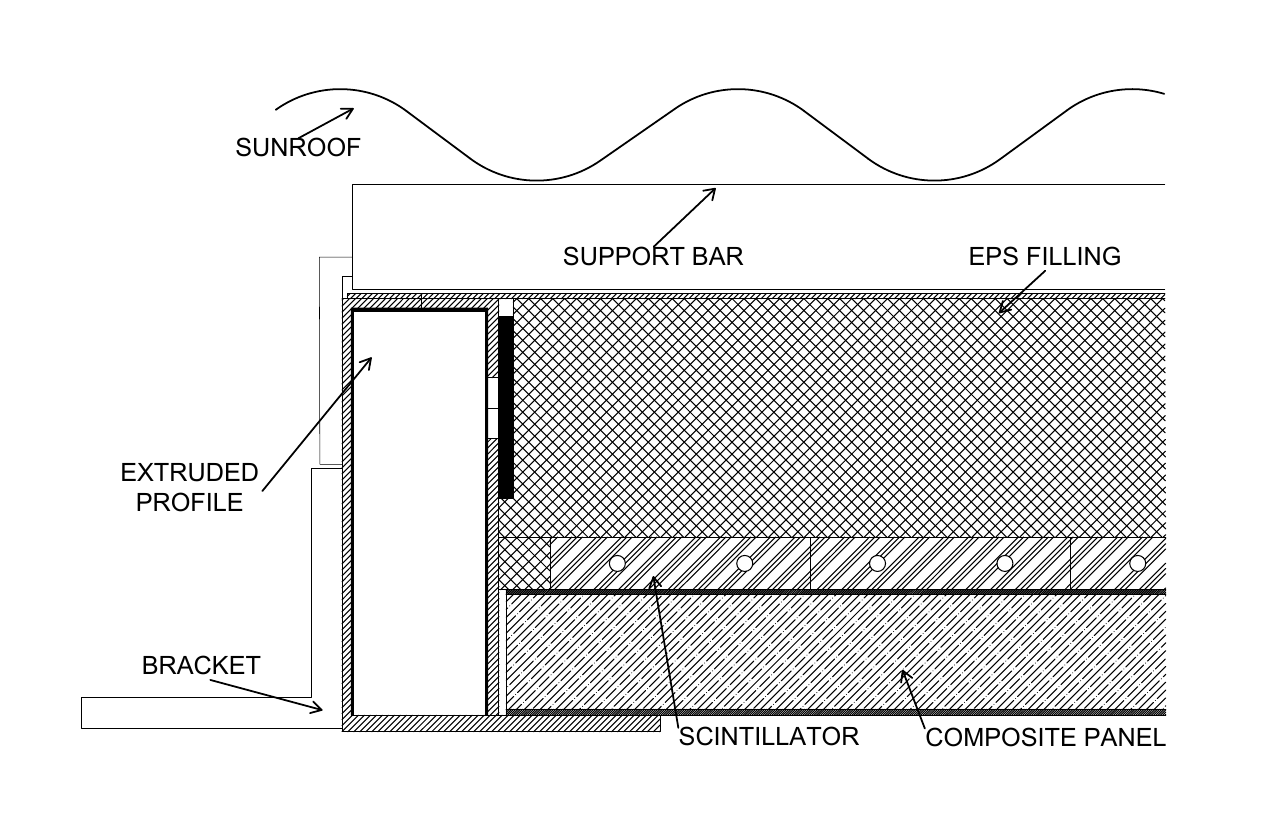}
\caption{Side-view detail of the SSD structure. The extruded profiles and the composite panel form the detector container. The scintillator bars (each with two lengthwise channels to hold optical fibers) are mounted on the composite panel. The brackets are used to mount the detector onto a support frame while the sunroof shades the SSD from direct sun light.}
\label{fig:sideview}
\end{figure}

The structure of the rectangular enclosure frame of the SSD consists of four hollow aluminium beams with a rectangular cross section of  $80\,\text{mm}\times30\,\text{mm}$ and a material thickness of 2\,mm; they are extruded with an L-shaped profile, forming a shelf in the bottom inner side of the enclosure. 
The structural integrity  is ensured by 
the bottom plate of the SSD, a panel with a ``sandwich'' structure consisting of a 1\,mm aluminium-layer bonded to each side of a 22\,mm extruded polystyrene (XPS) insulation layer lying on the bottom L-shaped profile shelf and sealed to it by means of the two-component silicone adhesive OttoColl S610~\cite{desglue} used for the multiple gluing processes in the assembly of the structure.  
The sealant must satisfy the needs for elasticity to account for movement and stress to the bond, excellent adhesion to guarantee light tightness, and resistance to UV, weathering, and ageing for the projected SSD lifetime of more than a decade. Several years of tests on detector prototypes in the field showed suitability of OttoColl S610. A side-view cross-section of the enclosure box is shown in \cref{fig:sideview}, where the main components are labeled.

\begin{figure}
\centering
\def\w{0.495}
\includegraphics[width=\w\textwidth]{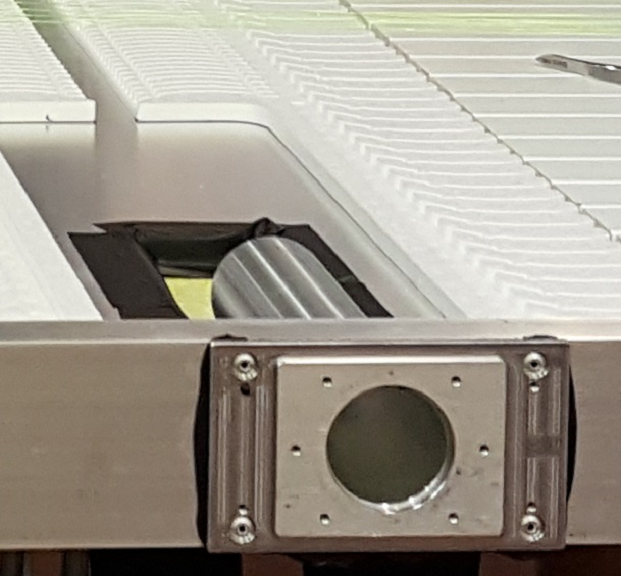}\hfill
\includegraphics[width=\w\textwidth]{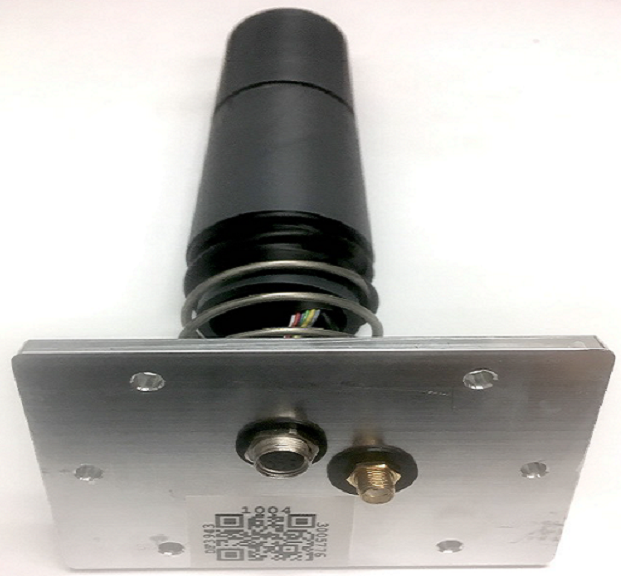}
\caption{\emph{Left:} Aluminium tube housing the readout device. The tube separates the electronics from the inner detector components and is fixed with glue and rivets onto the frame beam. \emph{Right:} The PVC tube containing the PMT unit. Visible are the spring to push the PMT to the end of the aluminium tube and the flange to close the aluminium tube.}
\label{fig:pmthouspvc}
\end{figure}

One of the long beams of the detector frame has a hole through which an aluminium tube for housing the PMT is inserted and fixed to the beam with glue and blind rivets (see \cref{fig:pmthouspvc}-Left).  The PMT, together with the High Voltage Power Supply and the accompanying electronics, integrated in the ISEG active base, is contained in an inner PVC tube (see \cref{fig:pmthouspvc}-Right). The PVC tube can be easily installed and removed from the aluminium tube, which is permanently fixed in the profile frame. A metal spring is used to push the PVC tube to the end of the aluminium tube, in order to ensure a close optical connection between the entrance glass window of the PMT and the exit of the collected light signal. The aluminium tube is closed from outside with a flange equipped with SubMiniature version A (SMA) and multipole connectors for an analog signal and a slow control cable, respectively. Additionally, a small aluminium box is mounted over the flange to protect cable connections to the PMT.

A  hole is drilled in the inner wall of the beam on the opposite side of the aluminium tube.
This hole is then covered with a porous but light-tight sintered metal plate. 
The sintered metal, together with multiple small holes on the bottom sides of the beam profiles, enables air exchange between the inner volume of the detector box and the outside. \
Air pressure equalization is necessary to prevent possible damage to the detector enclosure when the inner air volume expands due to increasing temperature, e.g.\ during daytime.

\begin{figure}
\centering
\def\w{0.7}
\includegraphics[width=\w\textwidth]{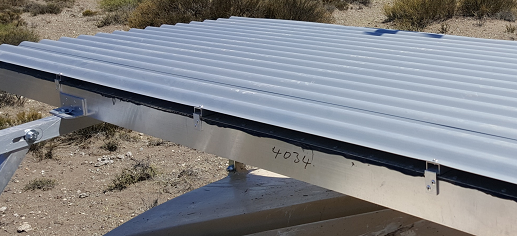}
\\[1mm]
\includegraphics[width=\w\textwidth]{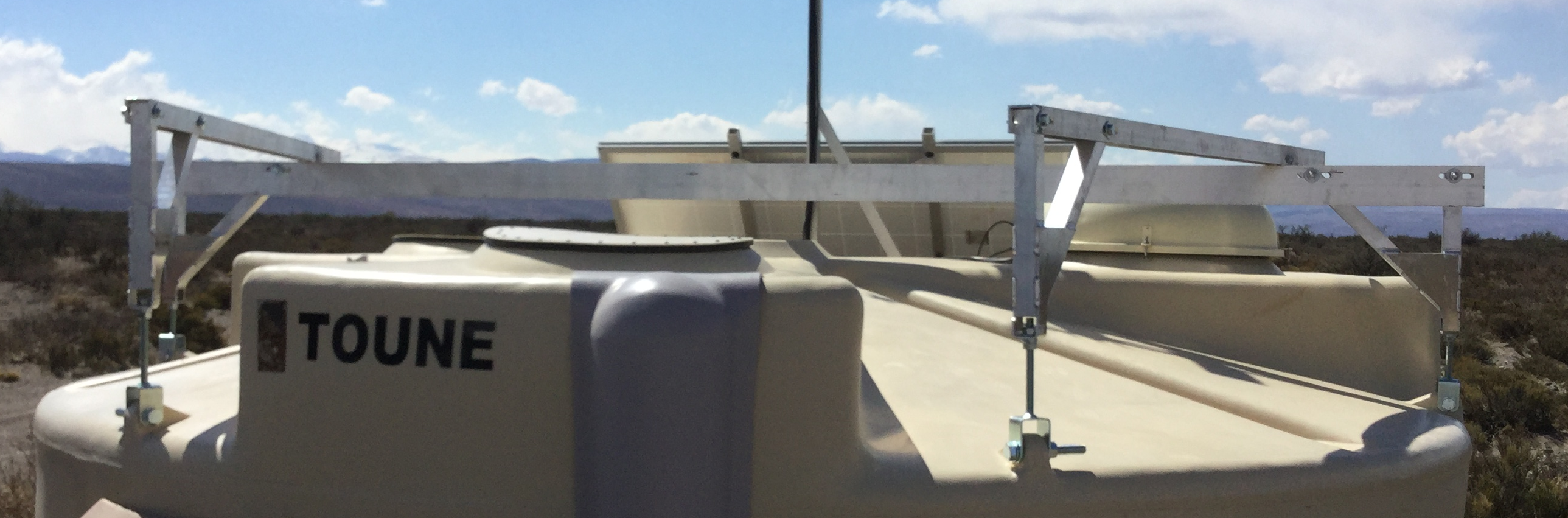}
\caption{\emph{Top:}
The sunroof of the SSD consisting of corrugated aluminium sheets. \emph{Bottom:} The SSD support structure on top of a WCD tank.}
\label{fig:topssd}
\end{figure}

The air volume inside the SSD module is reduced by filling the empty space in the module with expanded polystyrene (EPS)\@. 
The final closing of the SSDs is realized with a 1\,mm thick aluminium sheet which is glued to the frame profile to provide a light-tight seal. In addition, the seal is reinforced with closed-end blind rivets. 
To protect the SSDs from direct sun light, a sunroof consisting of corrugated aluminium sheets fixed to six aluminium bars is installed on top of each SSD, separated by 2\,cm, to allow air flow and reduce  temperature changes (see \cref{fig:topssd}-Top).

For the deployment in the field, four aluminium brackets are riveted to the outer side of the frame providing the connection of the SSD to the support frame.
The SSD detector units are mounted over the WCD tanks on a frame attached to four of the six lifting lugs molded into the tank. The  frame holds the SSD in the horizontal position and must be
adjustable enough to compensate for deformations of the plastic tanks standing on sandy ground for more than a decade. It is made from aluminium beams and can be easily assembled. Two shorter
beams are fixed to the main beam and this structure is supported by four legs (see \cref{fig:topssd}-Bottom).

The active part of each of the scintillator panels is composed of 24 scintillator elements (bars) of 1.6\,m length, 5\,cm width, and 1\,cm thickness.
The properties of plastic materials are affected by the temperature and temperature changes. Therefore, comprehensive aging studies of the scintillators and fibers were performed, yielding satisfactory results prior to the selection of the materials \cite{descifib}.

\begin{figure}
\centering
\def\w{0.96}
\includegraphics[width=\w\textwidth]{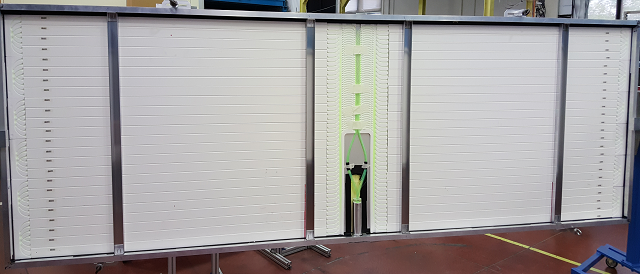}
\caption{ 
An open SSD where the scintillators have been secured with the tensioned aluminium bars.  The polystyrene routers and the fibers are also visible.}
\label{fig:ssdscint}
\end{figure}

The scintillator bars are extruded STYRON 663-W polystyrene, doped with 1\% of PPO (2,5-diphenyloxazole) and 0.03\% POPOP (1,4-bis(5-phenyl-2-oxazolyl)benzene), and are produced by the extrusion line of the Fermi National Accelerator Laboratory (FNAL)~\cite{Beznosko:2005ba}. Each bar is co-extruded with a 0.25\,mm-thick TiO$_2$ reflective layer, and with two lengthwise holes for wavelength-shifting fiber (WLS) insertion. 
The bars are fixed to the bottom plate via double-sided adhesive tape and firmly fixed to the external frame with two tensioned bars per panel.
The light produced in the scintillator bars is collected and propagated along the WLS fibers, Kuraray Y11(300)M S-type, inserted lengthwise in the bars and  positioned  with the help of polystyrene routers outside the bars (see \cref{fig:ssdscint}). 

At the outer ends of the scintillator panels, the routers form a ``U'' configuration of the WLS fibers with a curvature radius of 5\,cm, meaning that each fiber exiting from one hole of one scintillator bar enters the corresponding hole of another bar at a distance of 10\,cm (see \cref{fig:schema}).
Between the panels, the inner router grooves are designed to provide an equal path length for the fibers from the scintillator bars to the PMT. 
Each fiber has therefore the same length and is read out from both ends simultaneously to optimize the longitudinal uniformity of light response.

\begin{figure}
\centering
\def\h{0.24}
\includegraphics[height=\h\textheight]{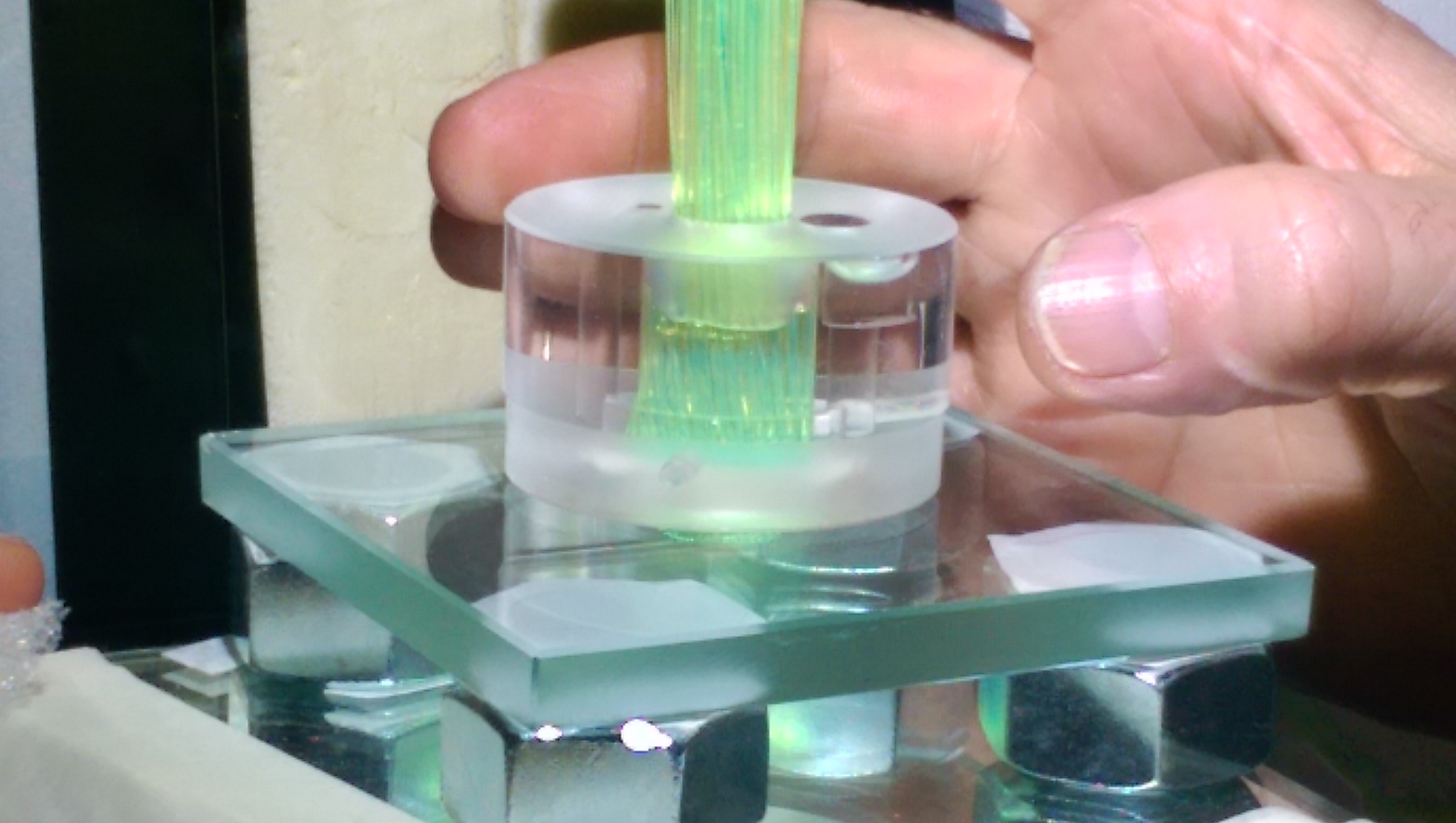}\hfill
\includegraphics[height=\h\textheight]{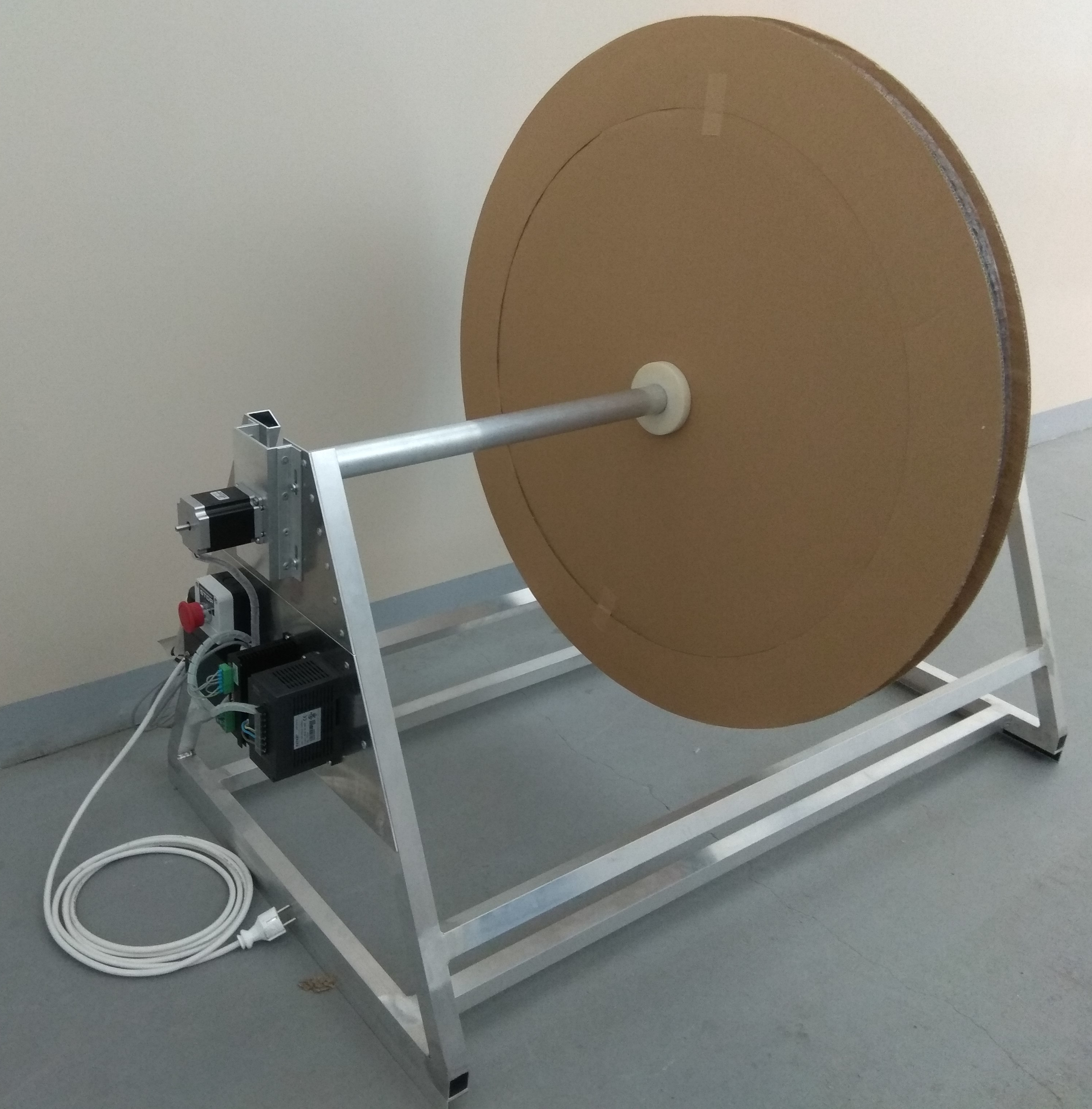}
\caption{\emph{Left:} 
The PMMA cylinder with the bundled and glued optical fibers.  \emph{Right:} Motorized device to unwind optical fiber from the roll.}
\label{fig:cookiedes}
\end{figure}

The fibers are bundled and glued with optical cement in a PMMA (poly(methyl methacrylate)) cylinder, a so-called ``cookie'' (\cref{fig:cookiedes}-Left) whose front window is connected to the PMT, a bi-alkali Hamamatsu R9420, 1.5-inch diameter, with 18\% quantum efficiency at a wavelength of 500\,nm. 

The PMMA material is transparent for photons with wavelength above 450\,nm and  transmission-related aging effects are negligible.
The cylinder has an outer diameter of 50\,mm, and a hole of 13\,mm diameter for inserting the fiber ends.
In addition, it contains two small holes for the gluing process, i.e.\ one hole for filling the glue and one hole for the escaping air.
The hole for the fiber ends is closed on the PMT side with a window consisting of a PMMA disk with a thickness of 6\,mm.
The window prevents leakage of any optical cement, and the fiber ends are protected from direct contact with the air or the glass of the photocathode.
Additionally, the window is slightly diffusive which results in a more uniform illumination of the photocathode.
The fiber ends are aligned inside the cylinder to the same distance of around 1mm to 2mm above the window, and glued using the two-component optical glue Eljen EJ-500. For a front view of the fiber bundle in the cookie, please see \cref{fig:cookie_photo}.


\section{Production of the SSD}

The complete production of SSDs was done within the Collaboration.
The task of producing more than 1500 SSD units exceeds the capabilities of any single laboratory available, so the work was distributed: the detectors were assembled and tested in parallel at multiple assembly facilities.
Many of the components were purchased or manufactured by one institution, and then distributed among partners based on demand.
The remaining components and materials were procured locally by each institution -- aluminium profiles for the frames were ordered from commercial manufacturers, and small mechanical parts were bought or produced by local workshops.
The assembly of the SSD detectors was distributed among six Auger institutions, with two more Auger laboratories working on preparing the PMT kits.
To ensure uniform production quality at all assembly sites, and based on experience with prototype detectors, a full Quality Production Plan was established and followed. This plan includes or references the detailed Assembly procedure, the Test \& Validation specifications and procedure, the tests results and conformity policy, as well as the packing, storage and shipment specifications.

All necessary equipment and tools were prepared at each assembly site, including tables with clamps appropriate for assembling SSD enclosures plus tiltable tables to prepare the optical "cookies" (see previous section).
The area in which scintillators and fibers were assembled was protected from sunlight and was equipped with non-UV lighting to prevent fiber degradation.
Also, a stand for optical fiber spools was needed, preferably equipped with an adjustable speed motor (\cref{fig:cookiedes}-Right).
Since the weight of an SSD module exceeds 100\,kg, a crane or other means to move the modules around was required.

\begin{figure}
\centering
\def\w{0.495}
\includegraphics[width=\w\textwidth]{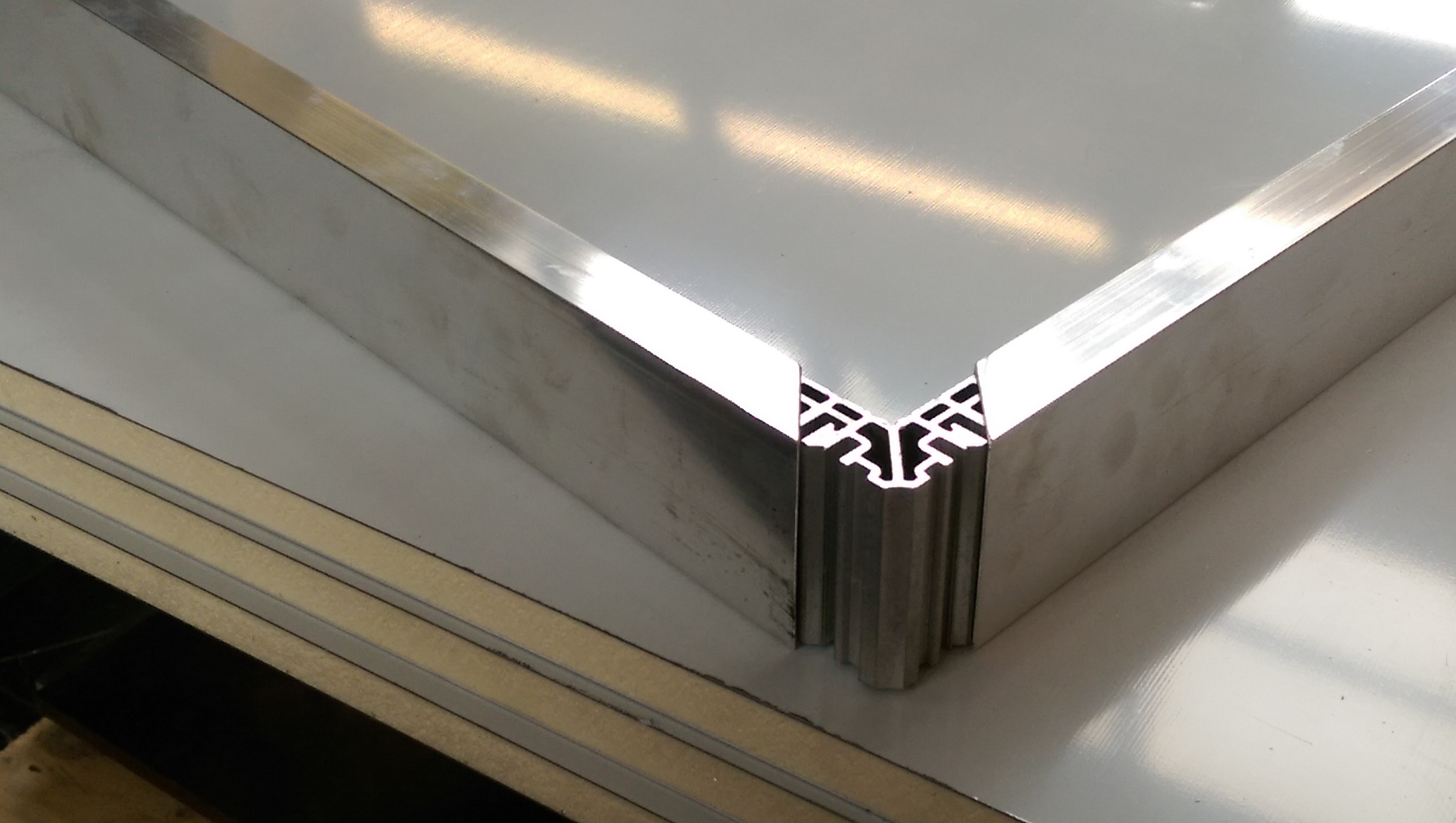}\hfill
\includegraphics[width=\w\textwidth]{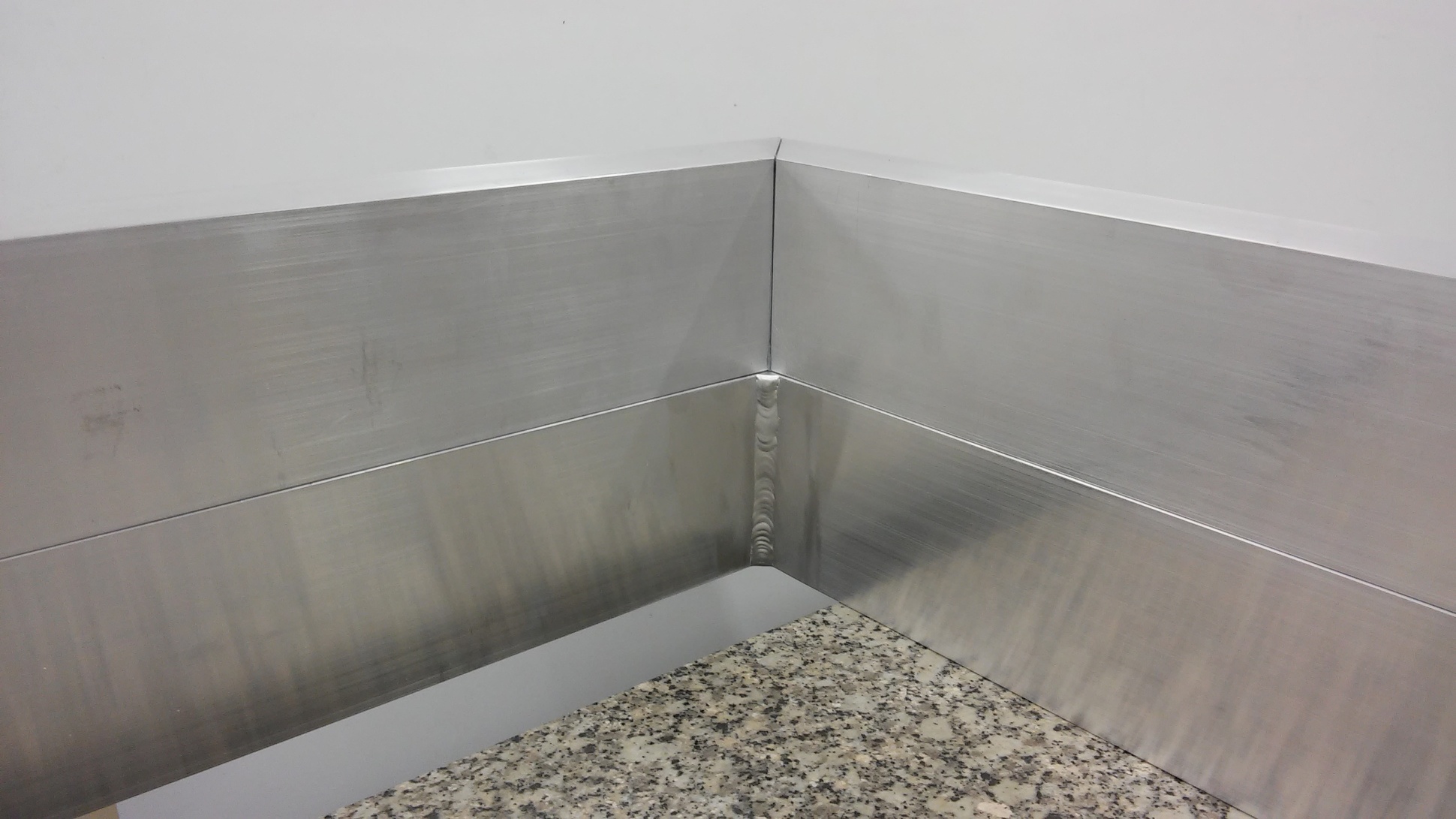}
\caption{\emph{Left:} Technique to make SSD frame corners by gluing extruded aluminium profiles to corner connectors. \emph{Right:} Prototype frame corners made using the gluing and welding techniques.}
\label{fig:corners}
\end{figure}

The assembly procedure starts with preparing the detector enclosure box: the frame is made of the extruded hollow  aluminium beams specified in the previous section.
The corners of the frame are made using corner connectors tightly fitting the hollow space of the beam (see \cref{fig:corners}-Left).
All elements are glued with OttoColl S610-Black silicone sealant.
Since it was possible to order a limited number (about 24\% of the total) of frames welded by a commercial supplier at an affordable price, welded frames were also used. Glued and welded frame corners are shown in \cref{fig:corners}-Right.

Prior to mounting the bottom composite panel into the frame, an appropriate cut-out is made in it, to allow placement of the aluminium tube housing the PMT.  Next, the panel is glued onto the frame. The aluminium tube and the sintered metal plate are also glued in. 
A good electrical connection (for grounding) between the bottom panel and the frame through the layer of the glue is ensured by additionally riveting the bottom panel to the frame. 

\begin{figure}
\centering
\def\w{0.495}
\includegraphics[width=\w\textwidth]{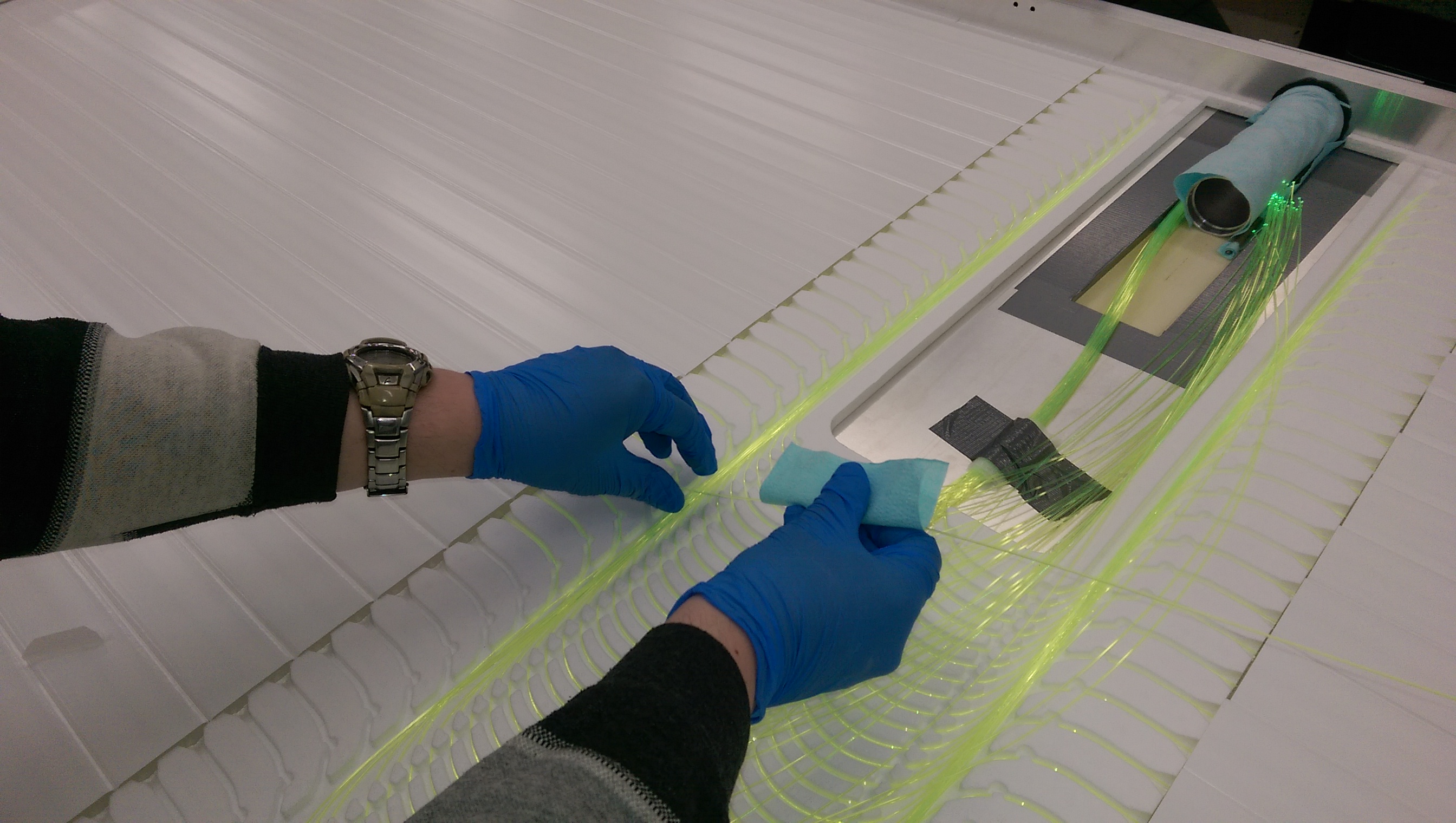}\hfill
\includegraphics[width=\w\textwidth]{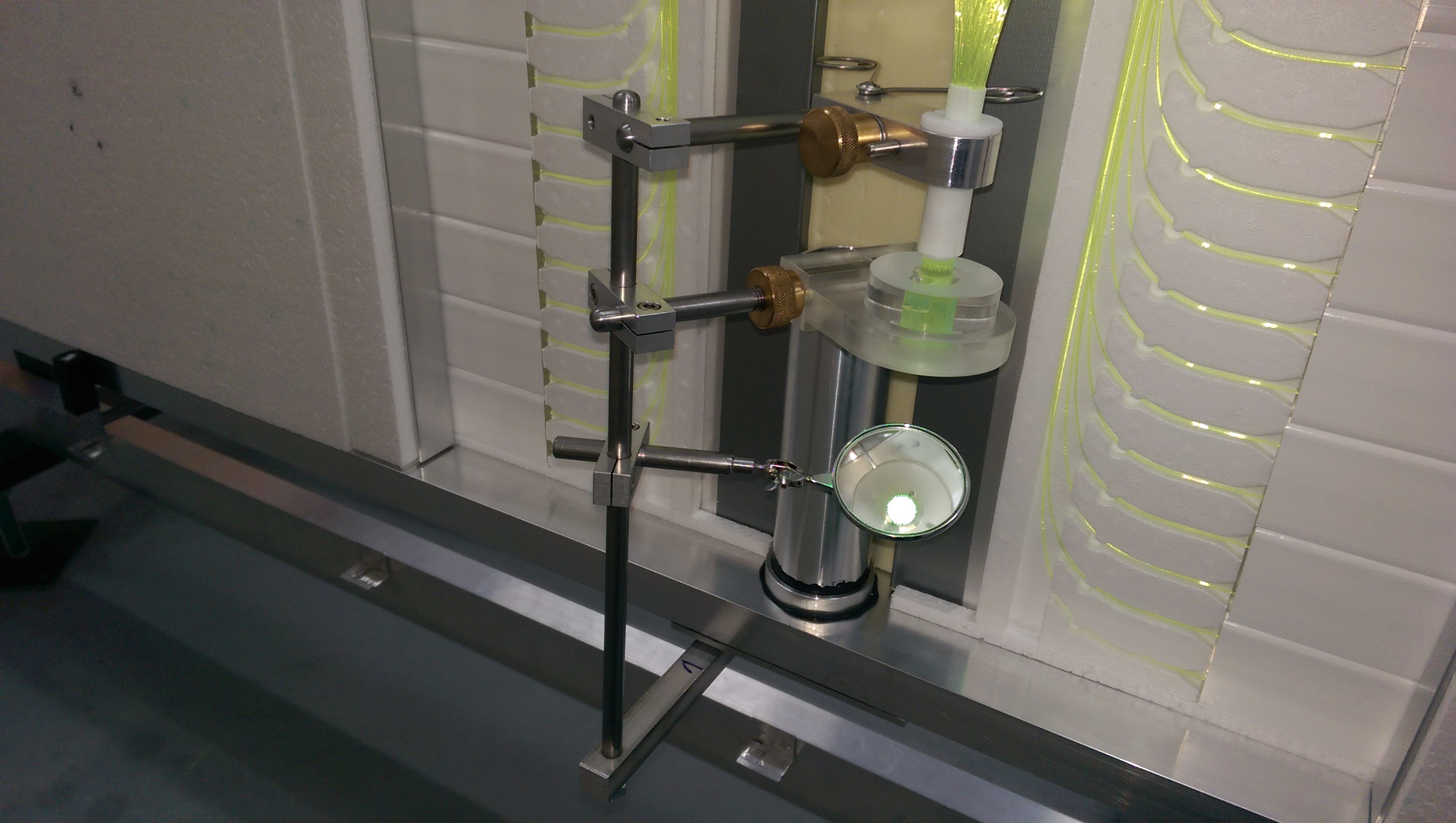}
\caption{\emph{Left:} Inserting fibers into scintillator bars. \emph{Right:} Example of a setup used for cookie preparation: fibers are aligned in the cookie and gluing quality is visually checked using a mirror.}
\label{fig:inserting}
\end{figure}

After the glue has cured, the scintillators are laid over the composite panel. Prior to that, the openings of channels that run through the scintillator bars are smoothed  
to get rid of sharp edges which might scratch the fibers.
Compressed air is then blown through the channels to make sure there are no small debris inside the channels that could prevent fiber insertion.
The outer fiber routers, the scintillator bars and the inner routers are then laid out inside the enclosure, and fixed in place with double-sided adhesive tape.
The optical fiber is unwound from the roll, cut into sections 5.85\,m long and inserted into the scintillator bars (\cref{fig:inserting}-Left) in a ``U'' configuration, as explained in the previous section.

All 96 fiber ends are bundled into the cookie.
With the module in a tilted position, the ends of the fibers are slightly melted by contacting them with a piece of borosilicate glass heated by an electric hotplate to around $140^\circ$C.
This reduces unwanted scattering of light on rough surfaces of fiber ends left after cutting.
The fiber ends are aligned at a distance of about 1.5\,mm from the cookie window to leave space for the optical glue.
After mixing the two components of the optical glue, it is necessary to remove air bubbles trapped within the liquid.
This is done by repeated degassing under vacuum (pressure ${\sim}0.1$ bar), with possible additional use of a centrifuge.
The cement is then poured into the cookie.
The cookie window is inspected from below using a mirror (\cref{fig:inserting}-Right). By tapping and rotating the cookie when needed, any bubbles remaining in front of the fiber ends are moved away from the path of light exiting the fibers.
Finally, a photograph of the finished cookie (\cref{fig:cookie_photo}) is taken and archived in a dedicated database.
After curing, the cookie is fixed to the PMT housing.
The space remaining within the enclosure is filled as designed and the top cover is glued and riveted to the frame.
With the mounting brackets and sunroof support bars fixed, the module is ready for testing (see \cref{sec:SSDtests}) and shipping to the Observatory. 

A total of 1519 SSD modules were produced. The sunroofs and protective aluminium boxes (to protect cable connection to the PMT) were added later at the Observatory, during deployment time.


\section{The PMT assembly and validation}

\begin{figure}
\centering
\includegraphics[width=0.7\textwidth]{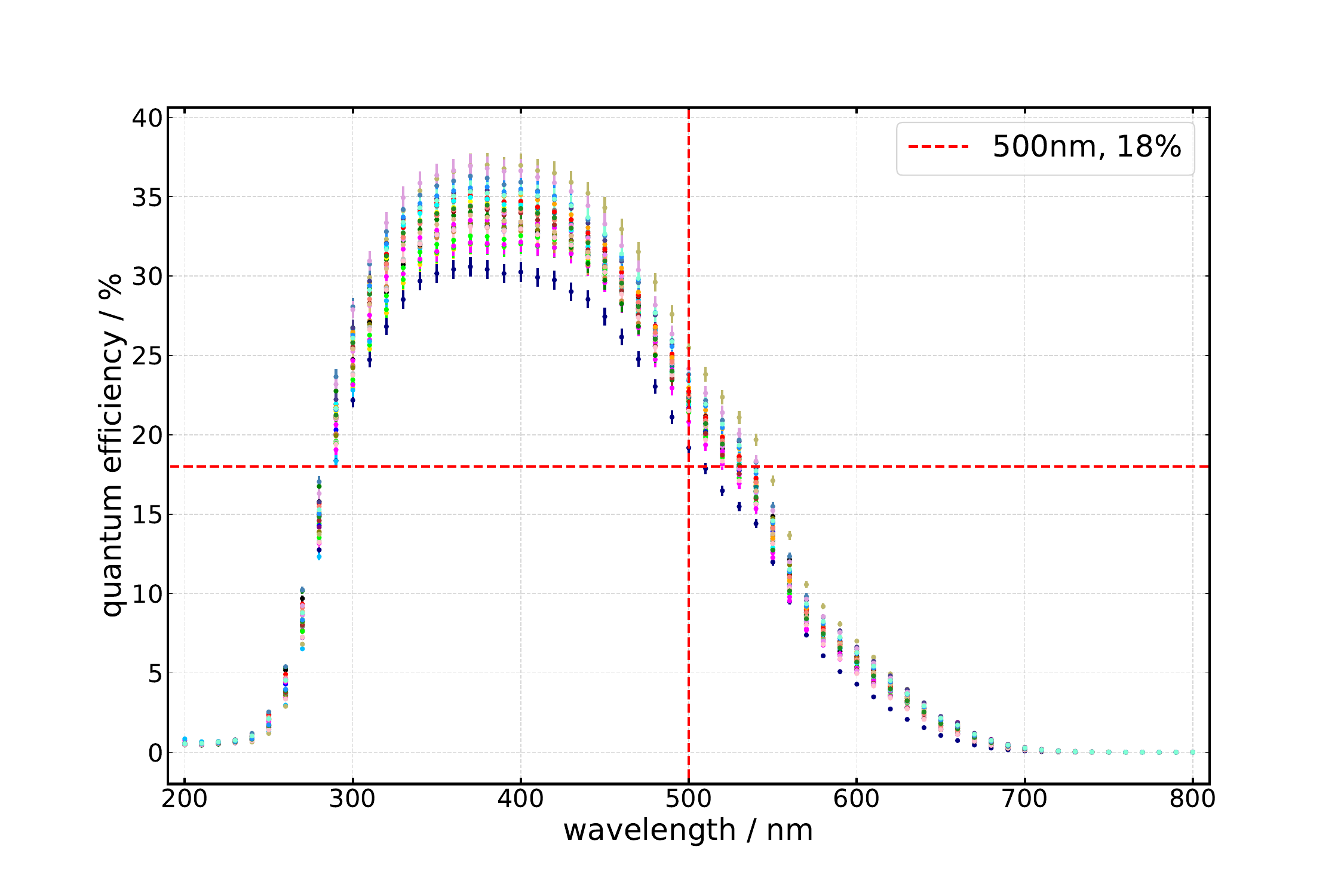}
\\
\includegraphics[width=0.7\textwidth]{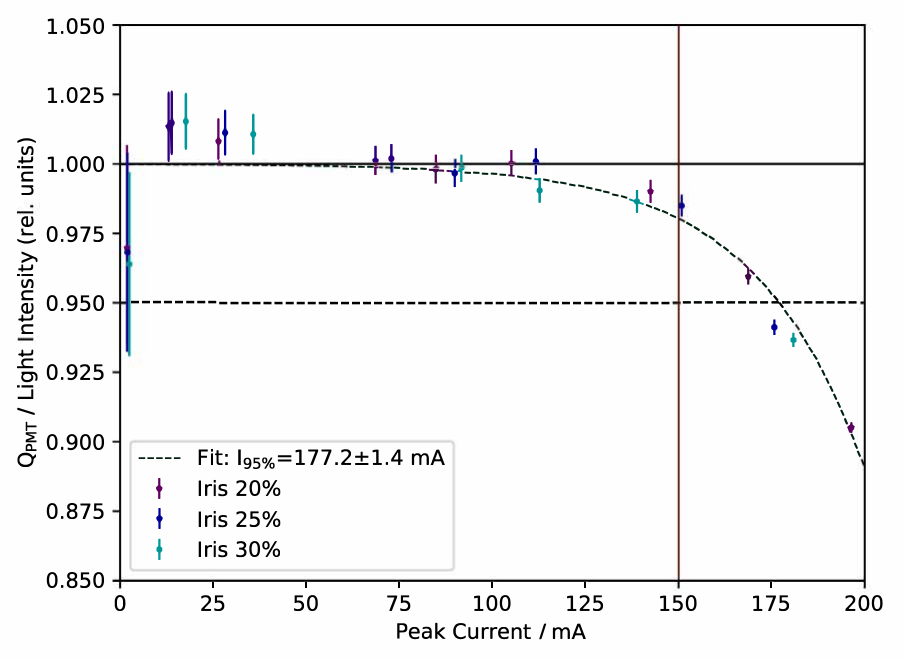}
\caption{\emph{Top:} Example of measured QE of several PMTs in the spectral range of interest where the lines indicate the specified minimum QE of 18\% at 500\,nm.  \emph{Bottom:} The linearity measured for one PMT operated at the nominal gain of $7\times 10^5$ with different openings of the iris diaphragm. For each iris setting, 10 different absorption filters set the incoming light intensity. The horizontal dashed line marks the maximum allowed deviation from linearity (95\%) and the vertical line marks the specification of 150 mA for the minimum linear peak current~\cite{Rautenberg:ICRC2021}.}
\label{fig:pmttests}
\end{figure}

The requirements for the light-detector used in the SSD are demanding.
It has to be very sensitive to register the signal of a minimum ionising particle (MIP) well above the noise level, 
and the signal is required to be linear for large signals up to 20\,000 MIP.  Several PMT candidates were evaluated for their suitability for the SSD detector, 
resulting in the selection of the Hamamatsu R9420.
The 1.5-inch PMT has an eight-stage dynode structure that allows for high linearity at a rather modest gain of $5.0{\times}10^4$.
Its quantum efficiency (QE) and its uniformity over the area of the photo-cathode have been measured.
In the region of the WLS fiber emission spectrum between 490\,nm and 550\,nm, its QE drops from 21\% to 15\%, as shown for a few PMTs in~\cref{fig:pmttests}-Top.
The photo-cathode uniformity was measured to be better than 3\%~\cite{Rautenberg:ICRC2021}.

For the cost- and power-efficient operation of the PMT, a base by ISEG is used, featuring an active voltage divider and integrated Cockcroft-Walton HV generation.
The ratio of the divider has been optimized for maximal linearity.

\begin{figure}
\centering
\includegraphics[width=0.8\textwidth]{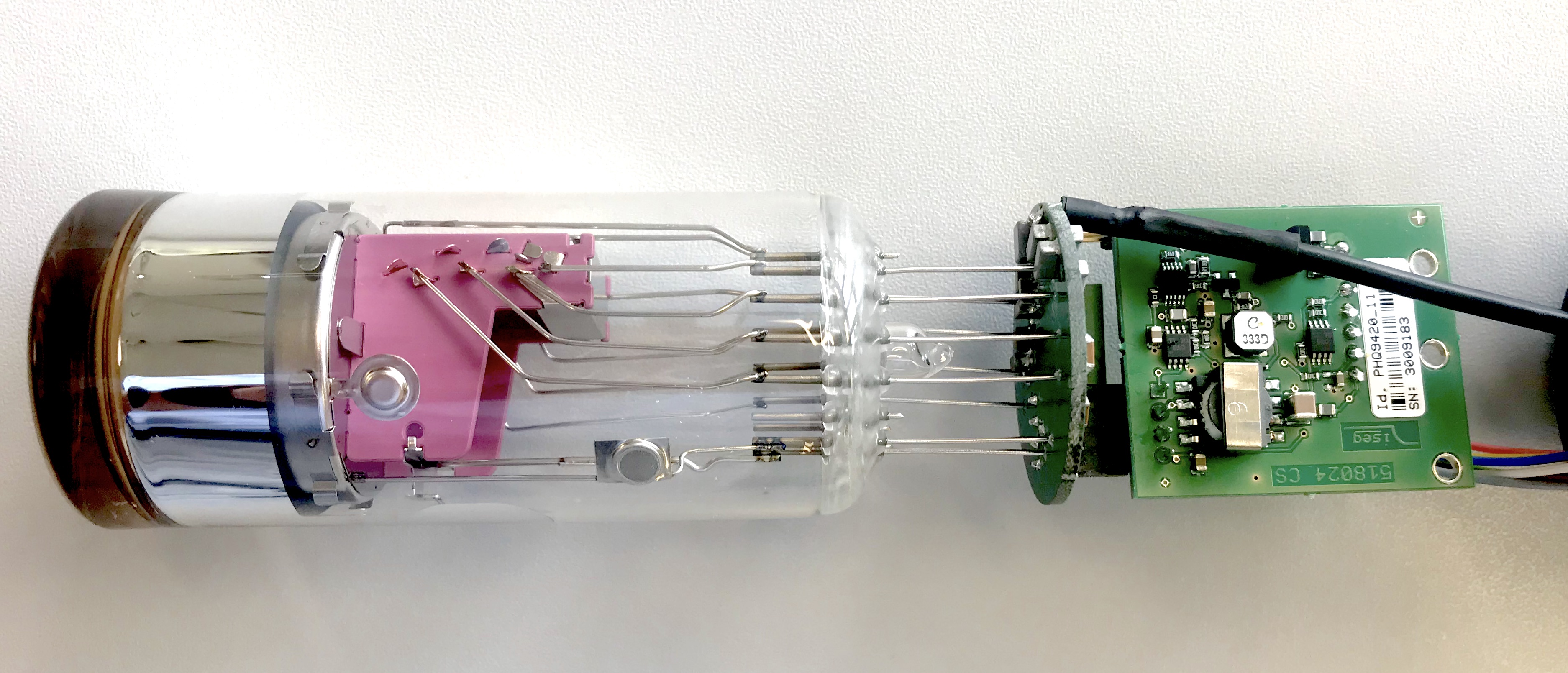}
\caption{Hamamatsu R9420 PMT soldered to a custom-designed ISEG base.}
\label{fig:pmtbase}
\end{figure}

In total, 1590 PMTs have been acquired, then assembled to PMT-units as shown in \cref{fig:pmthouspvc} by soldering the bases (see \cref{fig:pmtbase}), mounting the PMT in the PVC tube, and connecting the bases to the connectors on the flanges with the spring in-between.
For 10\% of the PMTs, the QE has been measured in the laboratory.
All PMT units have been tested in specialized setups~\cite{Longhitano_2018,Rautenberg:ICRC2021}.
Besides testing the general functionality, the linearity was measured at two gains - at the nominal gain of $7\times 10^5$ as well as at a gain of $5\times 10^4$, the operating gain in the experiment.
An example for the nominal gain to validate the specification of the producer is given in~\cref{fig:pmttests}-Bottom.
The full number of modules was delivered to the Observatory.

\section{SSD quality control and validation}
\label{sec:SSDtests}

Quality control and validation tests were systematically performed at each of the production sites for each SSD produced.
Although for practical reasons each site used slightly different equipment, common validation criteria and test procedure were followed at each site.
These included light-tightness, and several requirements related to the signal from vertical minimum ionizing particles (MIPs) - namely full efficiency and homogeneous response,  and a mean signal charge well above a qualification threshold of 15 single photo-electrons (SPE).

Constant attention to the quality of parts, with careful visual inspection during production, prevents including faulty or malfunctioning elements.
Any non-conformance observed during the assembly procedure or test phase was reported, documented and corrective actions were taken. 

\begin{figure}
\centering
\includegraphics[width=0.4\textwidth]{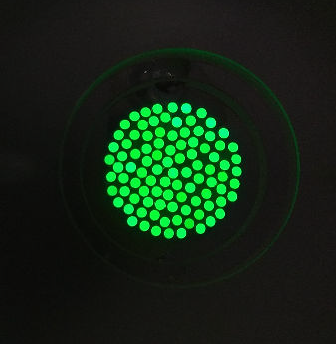}
\caption{Photograph of a cookie after the gluing of the fibers: the 96 ($2{\times}48$) fiber ends are seen through the cookie window and are shining uniformly.
The module is still open and uniformly illuminated by ambient light when this picture is taken.}
\label{fig:cookie_photo}
\end{figure}

Before closing the modules, the assembled ends of all the fibers are systematically inspected through the cookie window, first before and then after gluing them into the cookie.
A uniform illumination of all the module allows a check that all fibers ends are visible (no broken fibers or retracted fibers hidden by others, no air bubbles in front of fibers end).
A photograph of the cookie is taken and archived (\cref{fig:cookie_photo}). 

Once the cover is sealed and riveted, the module is moved to a test bench, equipped with a PMT and connected to a DAQ system.
The PMT/base used are of the same type as the final ones, but for the test the PMT is run at the larger nominal gain of $7\times 10^5$ compared to $5\times 10^4$ when operated in the field,
in order to make the single photo-electron calibration possible.
Some setups include additional signal amplification. 

The check for light leaks is carried out by monitoring the dark count rate under varying illumination conditions (blanket cover/uncover or day/night difference or by scanning the module using a bright light source). 

\begin{figure}
\centering
\def\h{0.32}
\includegraphics[height=\h\textwidth]{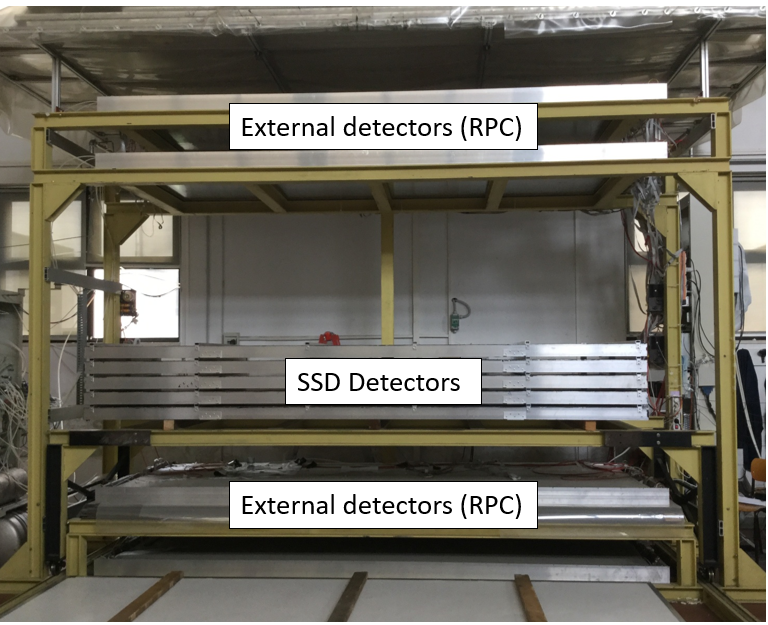}\hfill
\includegraphics[height=\h\textwidth]{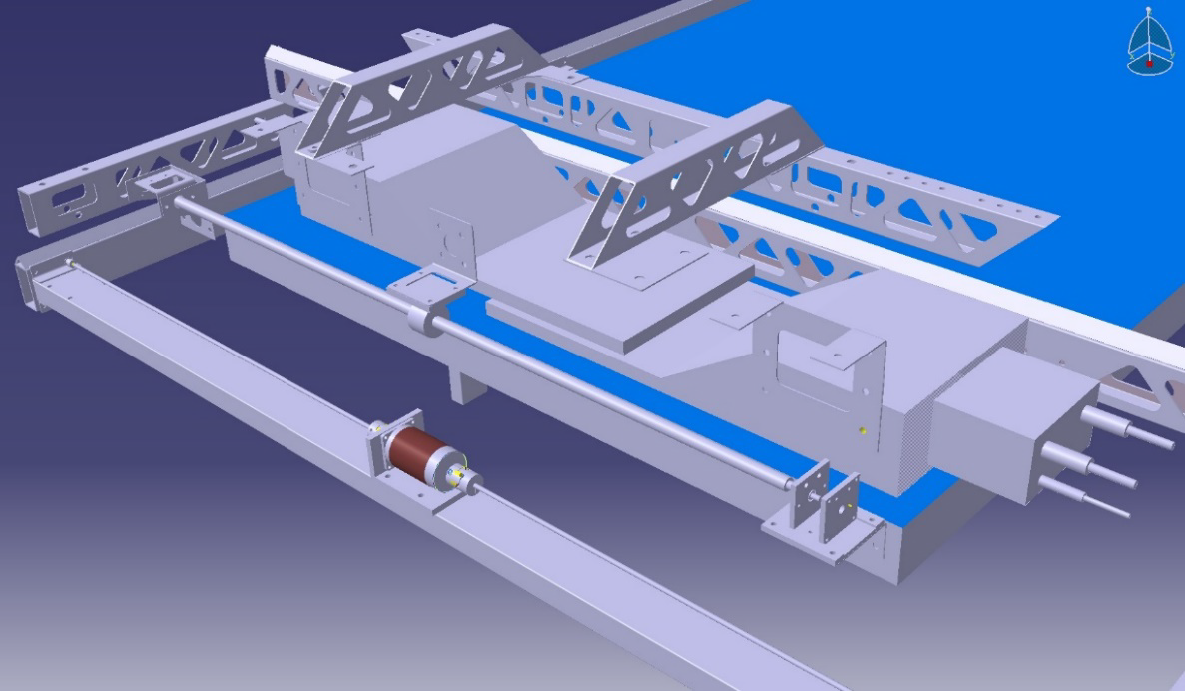}
\caption{\emph{Left:} One of the cosmic-ray test benches used.
\emph{Right:} A schematic drawing of the measurement setup consisting of an x-y scanner using 2 scintillator pads over the SSD (in blue).}
\label{fig:testbench_krakow}
\end{figure}

After the light-tightness is assessed, the detector response and efficiency are studied using cosmic-ray muons.
For this, different setups are employed at the different sites, making use of available equipment, ranging from a simple telescope made of plastic scintillators pads (\cref{fig:testbench_krakow}-Right) to full mapping with high granularity hodoscopes using streamer tubes or RPCs (\cref{fig:testbench_krakow}-Left).
A minimum requirement was to have two particle detectors working in coincidence, that would provide a trigger over a fraction of the SSD active area.
The measurement should cover as large and as representative a part of the SSD area as possible.
This was done by building the external detectors so they cover a fraction of all scintillators, on both sides of the SSD, during a measurement, or alternatively by moving smaller detectors over different parts of the SSD in the course of a test run. 
Independently of the setup used, the goal is to make sure that the module meets the required efficiency, uniformly across the two scintillator panels, and to characterize its response to single through-going muons. Data is taken over a period of about 12\,h triggering on external detectors.

\begin{figure}
\centering
\includegraphics[width=0.9\textwidth]{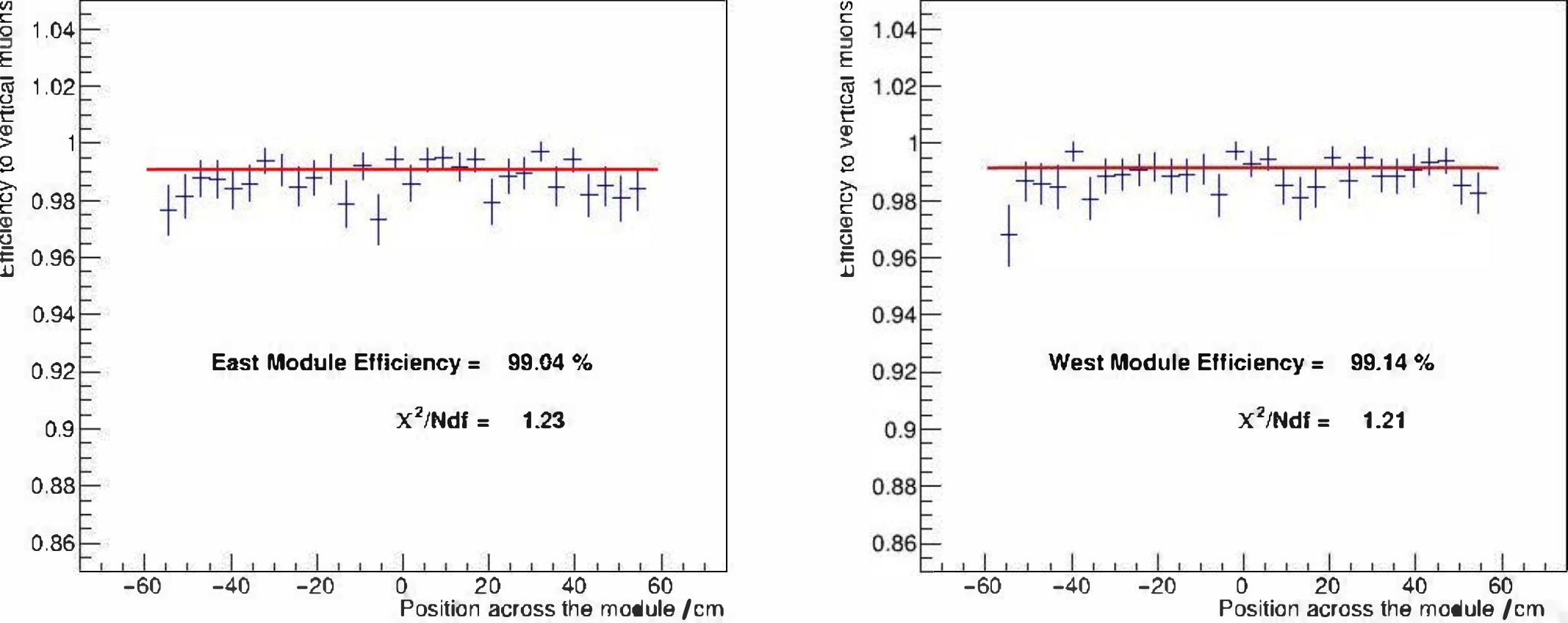}
\\[3mm]
\includegraphics[width=0.9\textwidth]{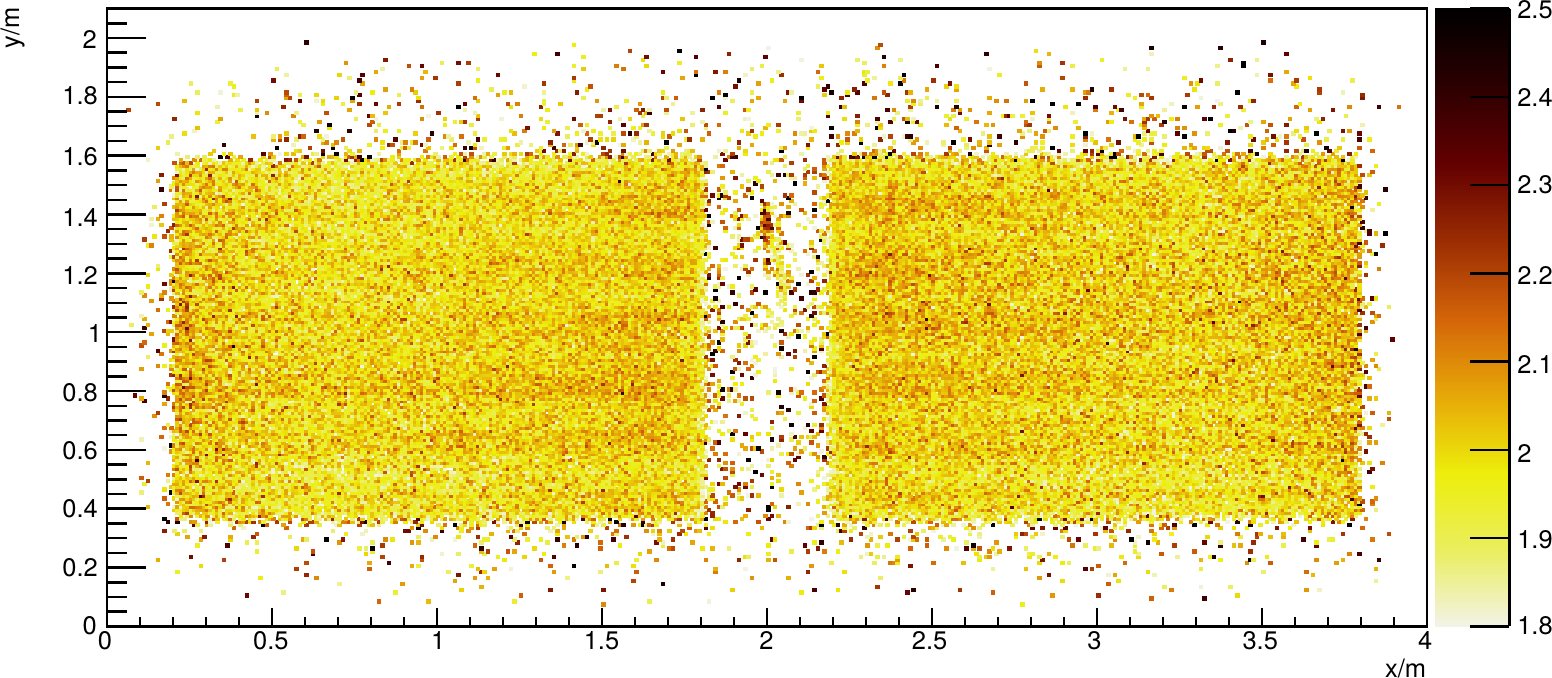}
\caption{Example of efficiency and uniformity measurements as a function of the position across the module.
\emph{Top:} efficiency of the two panels of one detector module.
\emph{Bottom:} uniformity of a detector, where the bundle of the fibers is also visible. The colour scale indicates the average decimal-logarithm of the deposited charge in pC. Note that the hodoscopes extend further out of the active parts of the SSD. Non-zero charges scattering outside of the active parts are actually induced by coincidental multiple muons events.}
\label{fig:positionmeas}
\end{figure}

The total efficiency of SSDs is measured in testing, as well as the uniformity of response over the active area of the detector; an example is shown in \cref{fig:positionmeas}.
Institutions that had the advantage of high-resolution detectors performed more detailed studies of some detectors.

These results showed that signals from individual scintillator bars deviate typically by no more than $10\%$ from the average. 
Signals from opposing ends of a scintillator bar differ by 
${\sim}5\%$, due to light attenuation in the fiber. 
For routine measurements at all sites, the shape of the MIP peak is used as an estimate of detector uniformity. This analysis is based on a histogram of all recorded signals obtained by triggering on quasi vertical muons. If a significant part of the detector were inefficient, it would cause a deformation of this peak.
A Gaussian fit sigma of the order of 30\% of the mean value  for quasi-vertical muons is a sign of satisfactory uniformity of the tested SSD.
Another test is done by measuring the trigger rate of each SSD.
Any significant deviation from the expected trigger rate (which should be stable for a fixed electronics setup) would be an indication of some problem with a given SSD.
A trigger rate smaller than expected could be caused by some damage in a part of the active area of the SSD or light loss in the cookie.
All test measurements are also stored in the SSD database, including e.g.\ values of MIP and SPE (charge in pC), histograms of measured signals, dark count rate, and a photograph of the cookie.

\begin{figure}
\centering
\includegraphics[width=0.525\textwidth]{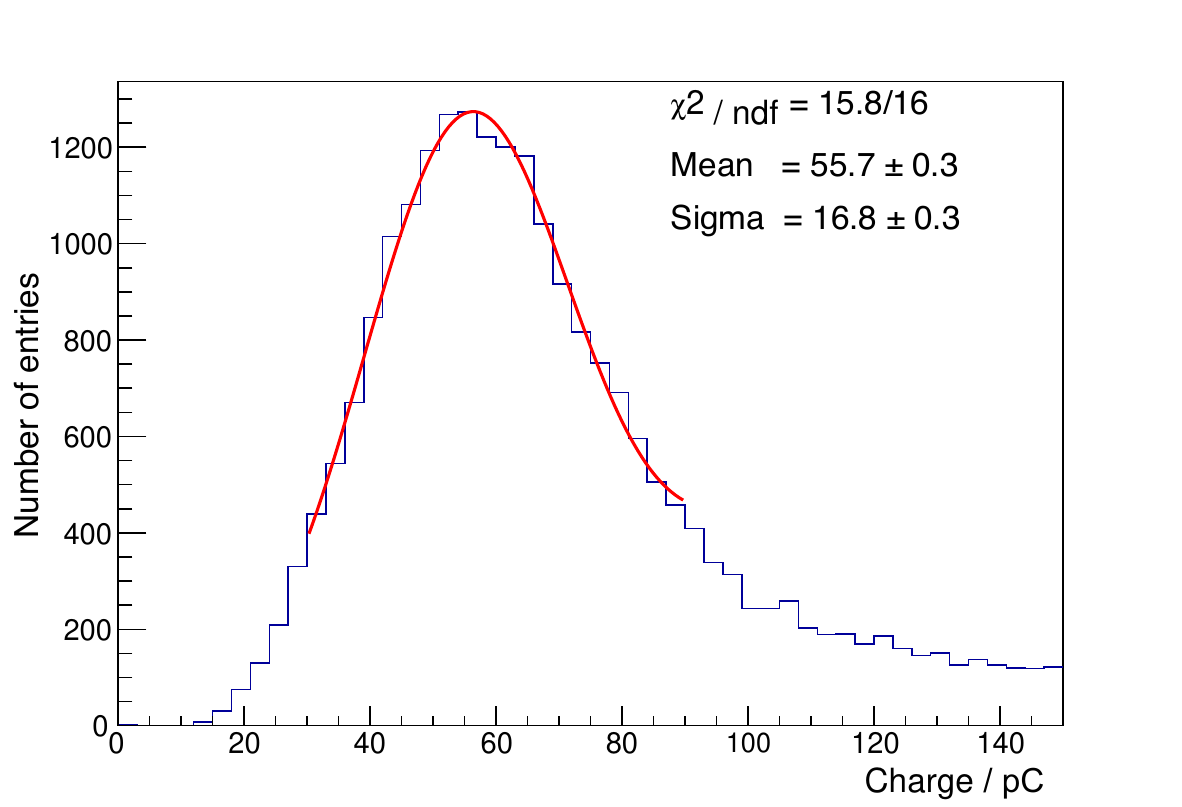}\hfill
\includegraphics[width=0.475\textwidth]{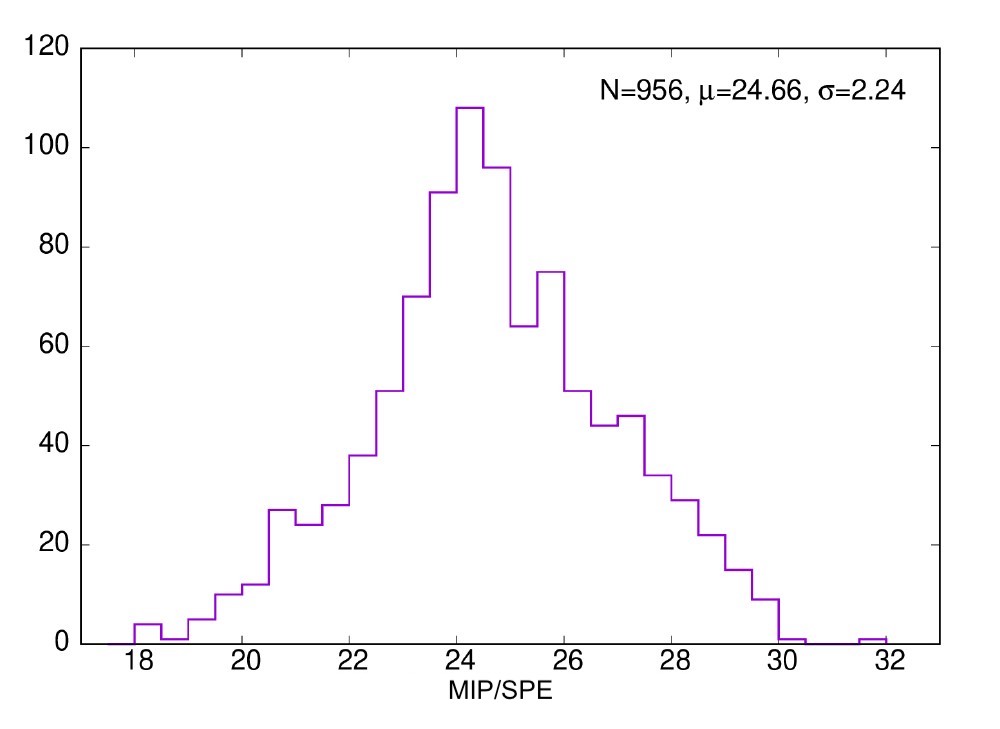}\hfill
\caption{\emph{Left:} Distribution of the charge values for quasi-vertical muons as measured for a single SSD module. The fit used here is a Gaussian distribution over a limited range to avoid a fraction of double hits (muon bundles or showers) in the distribution tail. Charges are in pC units. \emph{Right:} Distribution of average MIP value in units of SPE for a sample of the SSD detectors after rescaling.}
\label{fig:mipspeakfit}
\end{figure}

Fits of the peaks in the histograms of the signals produced by single muons are used to determine the mean charge of the MIP in pC. An example of the MIP distribution of a single SSD module and the corresponding fit is shown in \cref{fig:mipspeakfit}-Left. One should keep in mind that this histogram includes particles crossing a scintillator bar at wide range of inclination angles (approximately $\pm 20^\circ$).

The ratio of the MIP charge to the SPE charge can be used as an estimate of the quality of the tested SSD as it reflects the efficiency of the key detector components at generating (scintillators), collecting (fibers), and transmitting (cookie) the light.

Results from different institutions are obtained using  different PMTs and thus different quantum efficiencies.
To enable some comparison, 2 or 3 detectors from each production site were sent for test measurements to a single laboratory. 
These cross-tests are used to rescale the results obtained in other institutions.
The rescaled results show good agreement, with mean values of MIP/SPE ${\sim}25\pm2$, i.e.\ exceeding the design requirements by a factor of ${\sim}2$ (see \cref{fig:mipspeakfit}-Right).


\section{The SSD Final Assembly and Deployment}

Modules from the six production sites in Europe were shipped to the Auger site by sea freight in 40-feet maritime containers.
Each container could carry up to three pallets, totalling up to 51 modules per container. 
Once the containers were unloaded in the yard of the Observatory, the modules were visually inspected and stored outdoors in the yard.
Only in a few cases did the modules exhibit minor mechanical damage from handling.
Information including serial numbers and date of receipt was loaded into the inventory database. 
Support structures, corrugated metal sheets for sunroofs, and PMTs and accessories were shipped separately, and after reception were  stored for use in the assembly process. 
 
The final assembly of the SSD modules was performed in the main hall of the Assembly Building, which has an area of 240\,m$^2$ and is equipped with shelves for storage of components, three tables with wheels to hold and move one module each, and a bridge crane to handle the modules (approximately 120\,kg each). Modules are lifted using the brackets attached to the side of the frames. 
The assembly consisted of mounting the sunroof and the aluminium box which protects connections to the PMT.
The aluminium box was attached to the module with screws, and silicone was applied to ensure light and water tightness.
Support structures were also pre-assembled in the Assembly Building.
For this work, two technicians could assemble up to six modules and six support structures per day.

\begin{figure}
\centering
\includegraphics[width=0.7\textwidth]{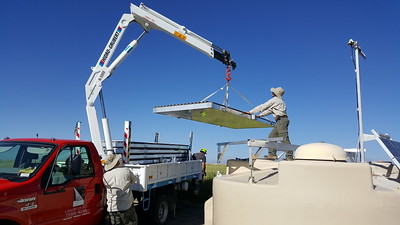}
\caption{Deployment of a SSD detector in the field.}
\label{fig:deploy}
\end{figure}

For the SSD deployment and installation in the field, a single truck was used.
It was a four-wheel drive Ford 4000 truck, with a crane mounted on the back and a loading area to transport the modules (\cref{fig:deploy}).
It could carry up to seven modules simultaneously.
In days of good weather, a team of three technicians could deploy typically 6 modules per day, requiring between 7 and 10 hours.
This includes the travel time to the installation region (on average, technicians had to travel a daily distance of 163\,km), and the time to move between detectors (typically, 1.5\,km of mostly off-road driving).
Deployment of SSD modules began in January 2019 and was completed in November 2021.
Most of the deployment activity was performed during the Covid-19 pandemic, which imposed additional limitations on the availability of staff for the activity.

The PMTs for the SSD were installed in the field at a later stage to avoid damaging them during the transport and handling of the modules. 
Mass deployment of SSD-PMTs was done by different teams, which also deployed new electronics for the surface detectors and installed a small PMT in the water-Cherenkov detectors.
This activity started in October 2021 and was completed in June 2023.

After installation, only a minimal number of SSD modules failed in the field.
In one of them, water had leaked inside the detector volume, also damaging the PMT.
Another one had a defective cookie and a third was replaced after it was exposed to a nearby bush fire which damaged its enclosure and might have damaged its interior. Overall, the installation of $1475$ SSDs has been very efficient and successful.


\section{Conclusions}

\begin{figure}
\centering
\def\h{0.4}
\includegraphics[height=\h\textwidth]{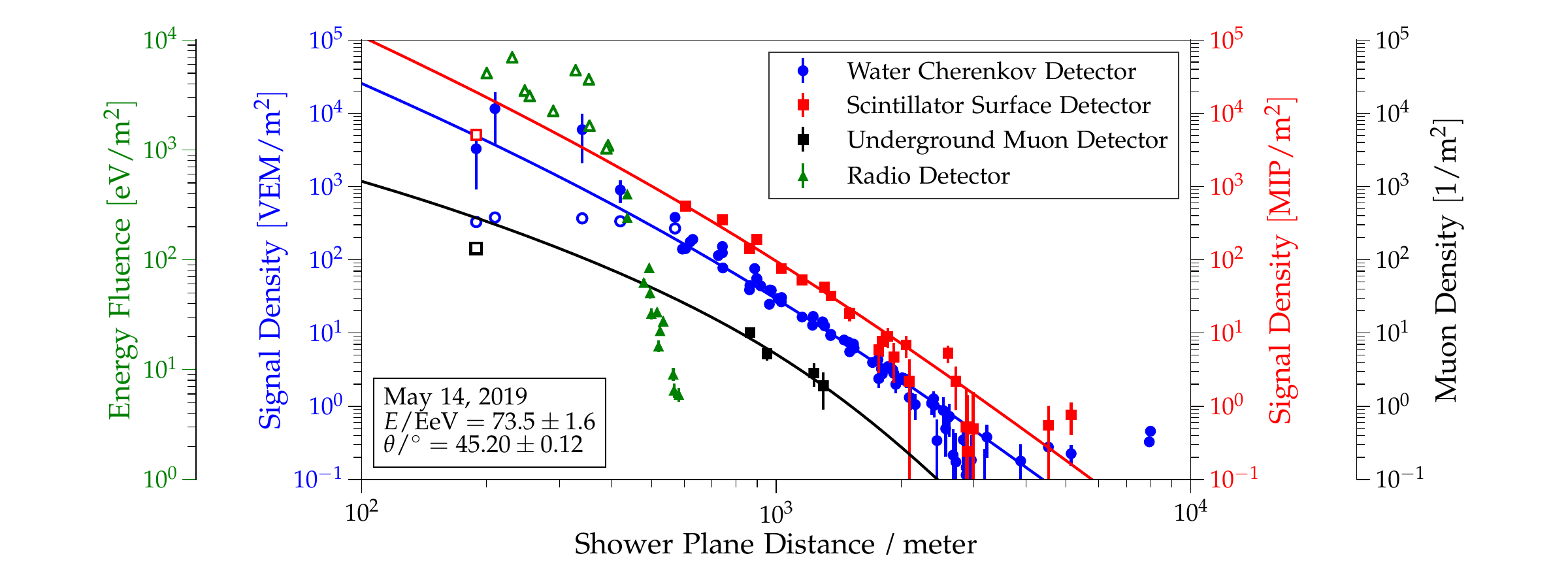}
\caption{Signal densities, as a function of the distance to the shower core, for one of the first high-energy events measured simultaneously with the SSD, WCD, UMD, and RD. The shower signal is measured in different observables according to specific features of a given detector: Minimum Ionizing Particles (MIP) for SSD; Vertical Equivalent Muons (VEM) for WCD; number of muons for UMD; electromagnetic wave  energy fluence for RD. Note that the distribution of the radio signal is considerably narrower than the distribution of particles in a shower.}
\label{fig:event}
\end{figure}

The upgraded Observatory, with its enhanced sensitivity to primary cosmic ray composition, afforded in part by the SSDs,
will play a fundamental role in the field of UHECRs for the next decade.
The new configuration of the Observatory will allow  multi-hybrid measurements of extensive air showers with a near-100\% duty cycle.
The showers will be simultaneously observed using water-Cherenkov detectors, scintillator detectors, radio detectors and underground muon counters, as shown in the real event displayed in \cref{fig:event}.

With these techniques, it will be possible to measure the muonic and electromagnetic components of the showers (and other characteristics) on an event-by-event basis and, therefore, help better infer the mass of the primary cosmic rays.
The improved determination of UHECR composition as a function of energy will help us understand the origin of the suppression in the cosmic ray spectrum and identify the regions of the nearby Universe where the UHECRs are accelerated.

\section*{Acknowledgments}

\begin{sloppypar}
The successful installation, commissioning, and operation of the Pierre
Auger Observatory would not have been possible without the strong
commitment and effort from the technical and administrative staff in
Malarg\"ue. We are very grateful to the following agencies and
organizations for financial support:
\end{sloppypar}

\begin{sloppypar}
Argentina -- Comisi\'on Nacional de Energ\'\i{}a At\'omica; Agencia Nacional de
Promoci\'on Cient\'\i{}fica y Tecnol\'ogica (ANPCyT); Consejo Nacional de
Investigaciones Cient\'\i{}ficas y T\'ecnicas (CONICET); Gobierno de la
Provincia de Mendoza; Municipalidad de Malarg\"ue; NDM Holdings and Valle
Las Le\~nas; in gratitude for their continuing cooperation over land
access; Australia -- the Australian Research Council; Belgium -- Fonds
de la Recherche Scientifique (FNRS); Research Foundation Flanders (FWO),
Marie Curie Action of the European Union Grant No.~101107047; Brazil --
Conselho Nacional de Desenvolvimento Cient\'\i{}fico e Tecnol\'ogico (CNPq);
Financiadora de Estudos e Projetos (FINEP); Funda\c{c}\~ao de Amparo \`a
Pesquisa do Estado de Rio de Janeiro (FAPERJ); S\~ao Paulo Research
Foundation (FAPESP) Grants No.~2019/10151-2, No.~2010/07359-6 and
No.~1999/05404-3; Minist\'erio da Ci\^encia, Tecnologia, Inova\c{c}\~oes e
Comunica\c{c}\~oes (MCTIC); Czech Republic -- GACR 24-13049S, CAS LQ100102401,
MEYS LM2023032, CZ.02.1.01/0.0/0.0/16{\textunderscore}013/0001402,
CZ.02.1.01/0.0/0.0/18{\textunderscore}046/0016010 and
CZ.02.1.01/0.0/0.0/17{\textunderscore}049/0008422 and CZ.02.01.01/00/22{\textunderscore}008/0004632;
France -- Centre de Calcul IN2P3/CNRS; Centre National de la Recherche
Scientifique (CNRS); Conseil R\'egional Ile-de-France; D\'epartement
Physique Nucl\'eaire et Corpusculaire (PNC-IN2P3/CNRS); D\'epartement
Sciences de l'Univers (SDU-INSU/CNRS); Institut Lagrange de Paris (ILP)
Grant No.~LABEX ANR-10-LABX-63 within the Investissements d'Avenir
Programme Grant No.~ANR-11-IDEX-0004-02; Germany -- Bundesministerium
f\"ur Bildung und Forschung (BMBF); Deutsche Forschungsgemeinschaft (DFG);
Finanzministerium Baden-W\"urttemberg; Helmholtz Alliance for
Astroparticle Physics (HAP); Helmholtz-Gemeinschaft Deutscher
Forschungszentren (HGF); Ministerium f\"ur Kultur und Wissenschaft des
Landes Nordrhein-Westfalen; Ministerium f\"ur Wissenschaft, Forschung und
Kunst des Landes Baden-W\"urttemberg; Italy -- Istituto Nazionale di
Fisica Nucleare (INFN); Istituto Nazionale di Astrofisica (INAF);
Ministero dell'Universit\`a e della Ricerca (MUR); CETEMPS Center of
Excellence; Ministero degli Affari Esteri (MAE), ICSC Centro Nazionale
di Ricerca in High Performance Computing, Big Data and Quantum
Computing, funded by European Union NextGenerationEU, reference code
CN{\textunderscore}00000013; M\'exico -- Consejo Nacional de Ciencia y Tecnolog\'\i{}a
(CONACYT) No.~167733; Universidad Nacional Aut\'onoma de M\'exico (UNAM);
PAPIIT DGAPA-UNAM; The Netherlands -- Ministry of Education, Culture and
Science; Netherlands Organisation for Scientific Research (NWO); Dutch
national e-infrastructure with the support of SURF Cooperative; Poland
-- Ministry of Science and Higher Education, grants No.~DIR/WK/2018/11 and
2022/WK/12; National Science Centre, grants No.~2016/22/M/ST9/00198,
2016/23/B/ST9/01635, 2020/39/B/ST9/01398, and 2022/45/B/ST9/02163;
Portugal -- Portuguese national funds and FEDER funds within Programa
Operacional Factores de Competitividade through Funda\c{c}\~ao para a Ci\^encia
e a Tecnologia (COMPETE); Romania -- Ministry of Research, Innovation
and Digitization, CNCS-UEFISCDI, contract no.~30N/2023 under Romanian
National Core Program LAPLAS VII, grant no.~PN 23 21 01 02 and project
number PN-III-P1-1.1-TE-2021-0924/TE57/2022, within PNCDI III; Slovenia
-- Slovenian Research Agency, grants P1-0031, P1-0385, I0-0033, N1-0111;
Spain -- Ministerio de Ciencia e Innovaci\'on/Agencia Estatal de
Investigaci\'on (PID2019-105544GB-I00, PID2022-140510NB-I00 and
RYC2019-027017-I), Xunta de Galicia (CIGUS Network of Research Centers,
Consolidaci\'on 2021 GRC GI-2033, ED431C-2021/22 and ED431F-2022/15),
Junta de Andaluc\'\i{}a (SOMM17/6104/UGR and P18-FR-4314), and the European
Union (Marie Sklodowska-Curie 101065027 and ERDF); USA -- Department of
Energy, Contracts No.~DE-AC02-07CH11359, No.~DE-FR02-04ER41300,
No.~DE-FG02-99ER41107 and No.~DE-SC0011689; National Science Foundation,
Grant No.~0450696, and NSF-2013199; The Grainger Foundation; Marie
Curie-IRSES/EPLANET; European Particle Physics Latin American Network;
and UNESCO.
\end{sloppypar}



\begin{center}
\rule{0.1\columnwidth}{0.5pt}\,\raisebox{-0.5pt}{\rule{0.05\columnwidth}{1.5pt}}~\raisebox{-0.375ex}{\scriptsize$\bullet$}~\raisebox{-0.5pt}{\rule{0.05\columnwidth}{1.5pt}}\,\rule{0.1\columnwidth}{0.5pt}
\end{center}

\section*{The Pierre Auger Collaboration}

{\small
A.~Abdul Halim$^{13}$,
P.~Abreu$^{70}$,
M.~Aglietta$^{53,51}$,
I.~Allekotte$^{1}$,
K.~Almeida Cheminant$^{78,77}$,
A.~Almela$^{7,12}$,
R.~Aloisio$^{44,45}$,
J.~Alvarez-Mu\~niz$^{76}$,
A.~Ambrosone$^{44}$,
J.~Ammerman Yebra$^{76}$
% ...
}
{\footnotesize
\begin{description}[labelsep=0.2em,align=right,labelwidth=0.7em,labelindent=0em,leftmargin=2em,noitemsep,before={\renewcommand\makelabel[1]{##1 }}]
\item[$^{1}$] Centro At\'omico Bariloche and Instituto Balseiro (CNEA-UNCuyo-CONICET), San Carlos de Bariloche, Argentina
\item[$^{2}$] Departamento de F\'\i{}sica and Departamento de Ciencias de la Atm\'osfera y los Oc\'eanos, FCEyN, Universidad de Buenos Aires and CONICET, Buenos Aires, Argentina
\item[$^{3}$] IFLP, Universidad Nacional de La Plata and CONICET, La Plata, Argentina
\item[$^{4}$] Instituto de Astronom\'\i{}a y F\'\i{}sica del Espacio (IAFE, CONICET-UBA), Buenos Aires, Argentina
\item[$^{5}$] Instituto de F\'\i{}sica de Rosario (IFIR) -- CONICET/U.N.R.\ and Facultad de Ciencias Bioqu\'\i{}micas y Farmac\'euticas U.N.R., Rosario, Argentina
\item[$^{6}$] Instituto de Tecnolog\'\i{}as en Detecci\'on y Astropart\'\i{}culas (CNEA, CONICET, UNSAM), and Universidad Tecnol\'ogica Nacional -- Facultad Regional Mendoza (CONICET/CNEA), Mendoza, Argentina
\item[$^{7}$] Instituto de Tecnolog\'\i{}as en Detecci\'on y Astropart\'\i{}culas (CNEA, CONICET, UNSAM), Buenos Aires, Argentina
\item[$^{8}$] International Center of Advanced Studies and Instituto de Ciencias F\'\i{}sicas, ECyT-UNSAM and CONICET, Campus Miguelete -- San Mart\'\i{}n, Buenos Aires, Argentina
\item[$^{9}$] Laboratorio Atm\'osfera -- Departamento de Investigaciones en L\'aseres y sus Aplicaciones -- UNIDEF (CITEDEF-CONICET), Argentina
\item[$^{10}$] Observatorio Pierre Auger, Malarg\"ue, Argentina
\item[$^{11}$] Observatorio Pierre Auger and Comisi\'on Nacional de Energ\'\i{}a At\'omica, Malarg\"ue, Argentina
\item[$^{12}$] Universidad Tecnol\'ogica Nacional -- Facultad Regional Buenos Aires, Buenos Aires, Argentina
\item[$^{13}$] University of Adelaide, Adelaide, S.A., Australia
\item[$^{14}$] Universit\'e Libre de Bruxelles (ULB), Brussels, Belgium
\item[$^{15}$] Vrije Universiteit Brussels, Brussels, Belgium
\item[$^{16}$] Centro Brasileiro de Pesquisas Fisicas, Rio de Janeiro, RJ, Brazil
\item[$^{17}$] Centro Federal de Educa\c{c}\~ao Tecnol\'ogica Celso Suckow da Fonseca, Petropolis, Brazil
\item[$^{18}$] Instituto Federal de Educa\c{c}\~ao, Ci\^encia e Tecnologia do Rio de Janeiro (IFRJ), Brazil
\item[$^{19}$] Universidade de S\~ao Paulo, Escola de Engenharia de Lorena, Lorena, SP, Brazil
\item[$^{20}$] Universidade de S\~ao Paulo, Instituto de F\'\i{}sica de S\~ao Carlos, S\~ao Carlos, SP, Brazil
\item[$^{21}$] Universidade de S\~ao Paulo, Instituto de F\'\i{}sica, S\~ao Paulo, SP, Brazil
\item[$^{22}$] Universidade Estadual de Campinas (UNICAMP), IFGW, Campinas, SP, Brazil
\item[$^{23}$] Universidade Estadual de Feira de Santana, Feira de Santana, Brazil
\item[$^{24}$] Universidade Federal de Campina Grande, Centro de Ciencias e Tecnologia, Campina Grande, Brazil
\item[$^{25}$] Universidade Federal do ABC, Santo Andr\'e, SP, Brazil
\item[$^{26}$] Universidade Federal do Paran\'a, Setor Palotina, Palotina, Brazil
\item[$^{27}$] Universidade Federal do Rio de Janeiro, Instituto de F\'\i{}sica, Rio de Janeiro, RJ, Brazil
\item[$^{28}$] Universidad de Medell\'\i{}n, Medell\'\i{}n, Colombia
\item[$^{29}$] Universidad Industrial de Santander, Bucaramanga, Colombia
\item[$^{30}$] Charles University, Faculty of Mathematics and Physics, Institute of Particle and Nuclear Physics, Prague, Czech Republic
\item[$^{31}$] Institute of Physics of the Czech Academy of Sciences, Prague, Czech Republic
\item[$^{32}$] Palacky University, Olomouc, Czech Republic
\item[$^{33}$] CNRS/IN2P3, IJCLab, Universit\'e Paris-Saclay, Orsay, France
\item[$^{34}$] Laboratoire de Physique Nucl\'eaire et de Hautes Energies (LPNHE), Sorbonne Universit\'e, Universit\'e de Paris, CNRS-IN2P3, Paris, France
\item[$^{35}$] Univ.\ Grenoble Alpes, CNRS, Grenoble Institute of Engineering Univ.\ Grenoble Alpes, LPSC-IN2P3, 38000 Grenoble, France
\item[$^{36}$] Universit\'e Paris-Saclay, CNRS/IN2P3, IJCLab, Orsay, France
\item[$^{37}$] Bergische Universit\"at Wuppertal, Department of Physics, Wuppertal, Germany
\item[$^{38}$] Karlsruhe Institute of Technology (KIT), Institute for Experimental Particle Physics, Karlsruhe, Germany
\item[$^{39}$] Karlsruhe Institute of Technology (KIT), Institut f\"ur Prozessdatenverarbeitung und Elektronik, Karlsruhe, Germany
\item[$^{40}$] Karlsruhe Institute of Technology (KIT), Institute for Astroparticle Physics, Karlsruhe, Germany
\item[$^{41}$] RWTH Aachen University, III.\ Physikalisches Institut A, Aachen, Germany
\item[$^{42}$] Universit\"at Hamburg, II.\ Institut f\"ur Theoretische Physik, Hamburg, Germany
\item[$^{43}$] Universit\"at Siegen, Department Physik -- Experimentelle Teilchenphysik, Siegen, Germany
\item[$^{44}$] Gran Sasso Science Institute, L'Aquila, Italy
\item[$^{45}$] INFN Laboratori Nazionali del Gran Sasso, Assergi (L'Aquila), Italy
\item[$^{46}$] INFN, Sezione di Catania, Catania, Italy
\item[$^{47}$] INFN, Sezione di Lecce, Lecce, Italy
\item[$^{48}$] INFN, Sezione di Milano, Milano, Italy
\item[$^{49}$] INFN, Sezione di Napoli, Napoli, Italy
\item[$^{50}$] INFN, Sezione di Roma ``Tor Vergata'', Roma, Italy
\item[$^{51}$] INFN, Sezione di Torino, Torino, Italy
\item[$^{52}$] Istituto di Astrofisica Spaziale e Fisica Cosmica di Palermo (INAF), Palermo, Italy
\item[$^{53}$] Osservatorio Astrofisico di Torino (INAF), Torino, Italy
\item[$^{54}$] Politecnico di Milano, Dipartimento di Scienze e Tecnologie Aerospaziali , Milano, Italy
\item[$^{55}$] Universit\`a del Salento, Dipartimento di Matematica e Fisica ``E.\ De Giorgi'', Lecce, Italy
\item[$^{56}$] Universit\`a dell'Aquila, Dipartimento di Scienze Fisiche e Chimiche, L'Aquila, Italy
\item[$^{57}$] Universit\`a di Catania, Dipartimento di Fisica e Astronomia ``Ettore Majorana``, Catania, Italy
\item[$^{58}$] Universit\`a di Milano, Dipartimento di Fisica, Milano, Italy
\item[$^{59}$] Universit\`a di Napoli ``Federico II'', Dipartimento di Fisica ``Ettore Pancini'', Napoli, Italy
\item[$^{60}$] Universit\`a di Palermo, Dipartimento di Fisica e Chimica ''E.\ Segr\`e'', Palermo, Italy
\item[$^{61}$] Universit\`a di Roma ``Tor Vergata'', Dipartimento di Fisica, Roma, Italy
\item[$^{62}$] Universit\`a Torino, Dipartimento di Fisica, Torino, Italy
\item[$^{63}$] Benem\'erita Universidad Aut\'onoma de Puebla, Puebla, M\'exico
\item[$^{64}$] Unidad Profesional Interdisciplinaria en Ingenier\'\i{}a y Tecnolog\'\i{}as Avanzadas del Instituto Polit\'ecnico Nacional (UPIITA-IPN), M\'exico, D.F., M\'exico
\item[$^{65}$] Universidad Aut\'onoma de Chiapas, Tuxtla Guti\'errez, Chiapas, M\'exico
\item[$^{66}$] Universidad Michoacana de San Nicol\'as de Hidalgo, Morelia, Michoac\'an, M\'exico
\item[$^{67}$] Universidad Nacional Aut\'onoma de M\'exico, M\'exico, D.F., M\'exico
\item[$^{68}$] Institute of Nuclear Physics PAN, Krakow, Poland
\item[$^{69}$] University of \L{}\'od\'z, Faculty of High-Energy Astrophysics,\L{}\'od\'z, Poland
\item[$^{70}$] Laborat\'orio de Instrumenta\c{c}\~ao e F\'\i{}sica Experimental de Part\'\i{}culas -- LIP and Instituto Superior T\'ecnico -- IST, Universidade de Lisboa -- UL, Lisboa, Portugal
\item[$^{71}$] ``Horia Hulubei'' National Institute for Physics and Nuclear Engineering, Bucharest-Magurele, Romania
\item[$^{72}$] Institute of Space Science, Bucharest-Magurele, Romania
\item[$^{73}$] Center for Astrophysics and Cosmology (CAC), University of Nova Gorica, Nova Gorica, Slovenia
\item[$^{74}$] Experimental Particle Physics Department, J.\ Stefan Institute, Ljubljana, Slovenia
\item[$^{75}$] Universidad de Granada and C.A.F.P.E., Granada, Spain
\item[$^{76}$] Instituto Galego de F\'\i{}sica de Altas Enerx\'\i{}as (IGFAE), Universidade de Santiago de Compostela, Santiago de Compostela, Spain
\item[$^{77}$] IMAPP, Radboud University Nijmegen, Nijmegen, The Netherlands
\item[$^{78}$] Nationaal Instituut voor Kernfysica en Hoge Energie Fysica (NIKHEF), Science Park, Amsterdam, The Netherlands
\item[$^{79}$] Stichting Astronomisch Onderzoek in Nederland (ASTRON), Dwingeloo, The Netherlands
\item[$^{80}$] Universiteit van Amsterdam, Faculty of Science, Amsterdam, The Netherlands
\item[$^{81}$] Case Western Reserve University, Cleveland, OH, USA
\item[$^{82}$] Colorado School of Mines, Golden, CO, USA
\item[$^{83}$] Department of Physics and Astronomy, Lehman College, City University of New York, Bronx, NY, USA
\item[$^{84}$] Michigan Technological University, Houghton, MI, USA
\item[$^{85}$] New York University, New York, NY, USA
\item[$^{86}$] University of Chicago, Enrico Fermi Institute, Chicago, IL, USA
\item[$^{87}$] University of Delaware, Department of Physics and Astronomy, Bartol Research Institute, Newark, DE, USA
\item[] -----
\item[$^{a}$] Max-Planck-Institut f\"ur Radioastronomie, Bonn, Germany
\item[$^{b}$] also at Kapteyn Institute, University of Groningen, Groningen, The Netherlands
\item[$^{c}$] now at GSI Helmholtzzentrum f\"ur Schwerionenforschung GmbH, Darmstadt, Germany
\item[$^{d}$] Fermi National Accelerator Laboratory, Fermilab, Batavia, IL, USA
\item[$^{e}$] Pennsylvania State University, University Park, PA, USA
\item[$^{f}$] Colorado State University, Fort Collins, CO, USA
\item[$^{g}$] Louisiana State University, Baton Rouge, LA, USA
\item[$^{h}$] now at Graduate School of Science, Osaka Metropolitan University, Osaka, Japan
\item[$^{i}$] Institut universitaire de France (IUF), France
\item[$^{j}$] now at Technische Universit\"at Dortmund and Ruhr-Universit\"at Bochum, Dortmund and Bochum, Germany
\end{description}
}

\end{document}